\def\year{2022}\relax
\documentclass[letterpaper]{article} 
\usepackage{aaai22}  
\usepackage{times}  
\usepackage{helvet}  
\usepackage{courier}  
\usepackage[hyphens]{url}  
\usepackage{graphicx} 
\usepackage{natbib}  
\usepackage{caption} 
\DeclareCaptionStyle{ruled}{labelfont=normalfont,labelsep=colon,strut=off} 
\usepackage{algorithm}
\usepackage{algorithmic}

\usepackage{newfloat}
\usepackage{listings}
\floatstyle{ruled}
\newfloat{listing}{tb}{lst}{}
\floatname{listing}{Listing}
\usepackage{booktabs}       
\usepackage{amsfonts}       
\usepackage{nicefrac}       
\usepackage{microtype}      
\usepackage[usenames]{color}
\usepackage[dvipsnames]{xcolor}         

\usepackage{amsthm}
\newtheorem{theorem}{Theorem}
\newtheorem{assumption}{Assumption}
\newtheorem{definition}{Definition}

\newtheorem{lemma}{Lemma}
\newtheorem{remark}{Remark}

\usepackage{amsmath,amssymb}       
\usepackage{mathbbol}
\usepackage{bm}
\usepackage{bbm}
\DeclareSymbolFontAlphabet{\amsmathbb}{AMSb}
\usepackage{mathtools}
\usepackage{accents}

\newcommand\munderbar[1]{%
  \underaccent{\bar}{#1}}

\usepackage{siunitx}

\DeclareMathOperator*{\minimize}{minimize}
\DeclareMathOperator*{\argmax}{arg\,max}

\usepackage[acronym]{glossaries}

\usepackage{fontawesome5}

\usepackage{xspace}
\usepackage{todonotes}
\usepackage{colonequals}
\usepackage[noadjust,nospace]{cite}
\usetikzlibrary{patterns,arrows,arrows.meta,calc,shapes,shadows,decorations.pathmorphing,decorations.pathreplacing,automata,shapes.multipart,positioning,shapes.geometric,fit,circuits,trees,shapes.gates.logic.US,fit, shadows.blur,shapes.symbols}

\usetikzlibrary{backgrounds,scopes}
\usepackage{marvosym}
\usepackage{multirow}

\usepackage{float}

\usepackage{tabularx}
\usepackage{booktabs}

\usepackage{caption}
\usepackage{subcaption}
\usepackage{appendix}
\usepackage{cleveref}

\crefname{equation}{Eq.}{Eq.}
\crefname{pluralequation}{Eqs.}{Eqs.}

\crefname{algorithm}{Algorithm}{Algorithm}

\crefname{figure}{Fig.}{Fig.}
\crefname{pluralfigure}{Figs.}{Figs.}

\crefname{section}{Sect.}{Sect.}
\crefname{pluralsection}{Sects.}{Sects.}

\crefname{table}{Table}{Table}
\crefname{pluraltable}{Tables}{Tables}

\crefname{definition}{Def.}{Def.}
\crefname{pluraldefinition}{Defs.}{Defs.}

\crefname{theorem}{Theorem}{Theorems}
\crefname{pluraltheorem}{Theorems}{Theorems}

\crefname{lemma}{Lemma}{Lemma}
\crefname{plurallemma}{Lemmas}{Lemmas}

\crefname{example}{Example}{Example}
\crefname{pluralexample}{Examples}{Examples}

\crefname{assumption}{Assumption}{Assumption}
\crefname{pluralassumption}{Assumptions}{Assumptions}

\crefname{remark}{Remark}{Remark}
\crefname{pluralremark}{Remarks}{Remarks}

\crefname{appendix}{Appendix}{Appendix}
\crefname{pluralappendix}{Appendices}{Appendices}

\usepackage{algorithmic}
\usepackage{tikz} 
\usepackage{pgfplots}
\usepgfplotslibrary{external,fillbetween}
\pgfplotsset{compat=1.8}

\definecolor{red}{rgb}{0.745,0.192,0.102}
\definecolor{darkgreen}{RGB}{34,161,55}
\definecolor{ruhuisstijlrood}{rgb}{0.745,0.192,0.102}
\definecolor{ruhuisstijlzwart}{rgb}{0,0,0}
\definecolor{ruhuisstijlwit}{rgb}{0.98,0.98,0.98}
\newcommand{\rured}[1]{\textcolor{ruhuisstijlrood}{#1}}
\newcommand{\darkgreen}[1]{\textcolor{darkgreen}{#1}}

\usepackage{mdframed}
\usepackage{enumitem}

\makeatletter
\def\orcidID#1{\smash{\href{http://orcid.org/#1}{\protect\raisebox{-1.25pt}{\protect\includegraphics{ORCID_Color.eps}}}}}
\makeatother

\usetikzlibrary{arrows,decorations.pathmorphing,positioning,fit,trees,shapes,shadows,automata,calc}

\newcommand{\prism}{\textrm{PRISM}\xspace}

\newcommand{\stochy}{\textrm{StocHy}\xspace}
\newcommand{\sreachtools}{\textrm{SReachTools}\xspace}
\newcommand{\probreach}{\textrm{ProbReach}\xspace}
\newcommand{\matlab}{\textrm{MATLAB}\xspace}

\newcommand{\R}{\amsmathbb{R}}

\newcommand{\N}{\amsmathbb{N}}
\newcommand{\distr}[1]{\ensuremath{\mathit{Dist(#1)}}}

\newcommand{\card}[1]{\ensuremath{\vert #1 \vert}}

\newcommand{\States}{S}
\newcommand{\goalStates}{\ensuremath{\States_{G}}}
\newcommand{\criticalStates}{\ensuremath{\States_{C}}}
\newcommand{\Actions}{Act}
\newcommand{\initState}{\ensuremath{s_I}}
\newcommand{\transfunc}{P}
\newcommand{\transfuncImdp}{\ensuremath{\mathcal{P}}}

\newcommand{\Interval}{\ensuremath{\amsmathbb{I}}}
\newcommand{\MDP}{\ensuremath{(\States,\Actions,\initState,\transfunc)}}
\newcommand{\iMDP}{\ensuremath{(\States,\Actions,\initState,\transfuncImdp)}}
\newcommand{\mdp}{\ensuremath{\mathcal{M}}}
\newcommand{\imdp}{\ensuremath{\mathcal{M}_\Interval}}

\newcommand{\policy}{\ensuremath{\pi}}
\newcommand{\policySpace}{\ensuremath{\Pi}}

\newcommand{\controller}{\ensuremath{\phi}}

\newcommand{\cStateSpace}{\ensuremath{\mathcal{X}}}
\newcommand{\cControlSpace}{\ensuremath{\mathcal{U}}}

\newcommand{\rank}[1]{\ensuremath{{\mathrm{rank}(#1)}}}

\newcommand{\Partition}{\ensuremath{\mathcal{R}}}
\newcommand{\goalRegion}{\ensuremath{\mathcal{X}_G}}
\newcommand{\criticalRegion}{\ensuremath{\mathcal{X}_C}}
\newcommand{\horizon}{\ensuremath{K}}

\newcommand{\Map}{\ensuremath{T}}

\newcommand{\property}{\ensuremath{\varphi_{\bm{x}_0}^\horizon}}
\newcommand{\propertyMDP}{\ensuremath{\varphi_{\initState}^\horizon}}

\newcommand{\reachProbCont}{\ensuremath{\mathit{Pr}^{\controller}( \property )}}

\newcommand{\reachProbDiscr}{\ensuremath{\mathit{Pr}^\policy(\propertyMDP)}}
\newcommand{\reachProbDiscrMin}{\ensuremath{\mathit{\underline{Pr}}^\policy(\propertyMDP)}}
\newcommand{\reachProbDiscrMinStar}{\ensuremath{\mathit{\underline{Pr}}^{\policy^*}(\propertyMDP)}}

\newcommand{\Gauss[2]}{\ensuremath{\mathcal{N}(#1 , #2)}}
\newcommand{\mean}{\ensuremath{\mu}}
\newcommand{\cov}{\ensuremath{\Sigma}}

\glsdisablehyper

\newacronym[]{LTL}{LTL}{linear temporal logic}
\newacronym[]{iid}{i.i.d.}{independent and identically distributed}
\newacronym[]{UAV}{UAV}{unmanned aerial vehicle}

\newacronym[]{PAC}{PAC}{probably approximately correct}
\newacronym[]{DRO}{DRO}{distributionally robust optimization}
\newacronym[plural=MDPs,firstplural=Markov decision processes (MDPs)]{MDP}{MDP}{Markov decision process}
\newacronym[plural=iMDPs,firstplural=interval Markov decision processes (iMDPs)]{iMDP}{iMDP}{interval Markov decision process}
\newacronym[plural=aMDPs,firstplural=augmented MDPs (aMDPs)]{aMDP}{aMDP}{augmented MDP}
\newacronym[plural=POMDPs,firstplural=partially observable Markov decision processes (POMDPs)]{POMDP}{POMDP}{partially observable Markov decision process}

\title{Sampling-Based Robust Control of Autonomous Systems \\ with Non-Gaussian Noise \thanks{
This work was funded by NWO grant NWA.1160.18.238 (PrimaVera), and ERC Advanced Grant 834115 (FUN2MODEL).
}}
\author{
    Thom S. Badings\textsuperscript{\rm 1}, Alessandro Abate\textsuperscript{\rm 2}, Nils Jansen\textsuperscript{\rm 1} \\
    David Parker\textsuperscript{\rm 3}, Hasan A. Poonawala\textsuperscript{\rm 4}, Marielle Stoelinga\textsuperscript{\rm 1}
}
\affiliations {
    \textsuperscript{\rm 1} Radboud University, Nijmegen, the Netherlands\\
    \textsuperscript{\rm 2} University of Oxford, Oxford, United Kingdom\\
    \textsuperscript{\rm 3} University of Birmingham, Birmingham, United Kingdom\\
    \textsuperscript{\rm 4} University of Kentucky, Kentucky, USA\\
}

\newif\ifappendix
\global\appendixtrue

\begin{document}

\maketitle

\begin{abstract}
Controllers for autonomous systems that operate in safety-critical settings must account for stochastic disturbances.
Such disturbances are often modeled as process noise, and common assumptions are that the underlying distributions are known and/or Gaussian.
In practice, however, these assumptions may be unrealistic and can lead to poor approximations of the true noise distribution.
We present a novel planning method that does not rely on any explicit representation of the noise distributions.
In particular, we address the problem of computing a controller that provides probabilistic guarantees on safely reaching a target.
First, we abstract the continuous system into a discrete-state model that captures noise by probabilistic transitions between states.
As a key contribution, we adapt tools from the scenario approach to compute probably approximately correct (PAC) bounds on these transition probabilities, based on a finite number of samples of the noise.
We capture these bounds in the transition probability intervals of a so-called interval Markov decision process (iMDP).
This iMDP is robust against uncertainty in the transition probabilities, and the tightness of the probability intervals can be controlled through the number of samples.
We use state-of-the-art verification techniques to provide guarantees on the iMDP, and compute a controller for which these guarantees carry over to the autonomous system.
Realistic benchmarks show the practical applicability of our method, even when the iMDP has millions of states or transitions.
\end{abstract}

\section{Introduction}

\label{sec:Introduction}

Consider a so-called \emph{reach-avoid problem} for an \gls{UAV}, where the goal is to reach a desirable region within a given time horizon, while avoiding certain unsafe regions~\citep{DBLP:books/daglib/BaierKatoen2008, DBLP:journals/toplas/ClarkeES86}.
A natural formal model for such an autonomous system is a \emph{dynamical system}.
The \emph{state} of the system reflects the position and velocity of the \gls{UAV}, and the \emph{control inputs} reflect choices that may change the state over time~\citep{Kulakowski2007dynamic}.
The dynamical system is \emph{linear} if the state transition is linear in the current state and control input.
Our problem is to compute a \emph{controller}, such that the state of the \gls{UAV} progresses safely, without entering unsafe regions, to its goal~\citep{aastrom2010feedback}.

However, factors like turbulence and wind gusts cause \emph{uncertainty} in the outcome of control inputs~\citep{DBLP:journals/trob/BlackmoreOBW10}.
We model such uncertainty as \emph{process noise}, which is an additive random variable (with possibly infinitely many outcomes) in the dynamical system that affects the transition of the state.
Controllers for autonomous systems that operate in safety-critical settings must account for such uncertainty.

A common assumption to achieve computational tractability of the problem is that the process noise follows a Gaussian distribution~\citep{DBLP:journals/spm/ParkSQ13}, for example in linear-quadratic-Gaussian control~\citep{anderson2007optimal}.
However, in realistic problems, such as the \gls{UAV} operating under turbulence, this assumption yields a poor approximation of the uncertainty~\citep{DBLP:journals/trob/BlackmoreOBW10}.
Distributions may even be \textit{unknown}, meaning that one cannot derive a set-bounded or stochastic representation of the noise.
In this case, it is generally hard or even impossible to derive \emph{hard guarantees} on the probability that a given controller ensures a safe progression of the system's state to the objective.

In this work, we do not require that the process noise is known.
Specifically, we provide \emph{\gls{PAC}} guarantees on the performance of a controller for the reach-avoid problem, where the distribution of the noise is unknown.
As such, we solve the following problem:

\begin{mdframed}[backgroundcolor=gray!20, nobreak=true, innerleftmargin=8pt, innerrightmargin=8pt]
Given a linear dynamical system perturbed by additive noise of unknown distribution, compute a controller under which, with high confidence, the probability to satisfy a reach-avoid problem is above a given threshold value.
\end{mdframed}

\subsubsection{Finite-state abstraction.}
The fundamental concept of our approach is to compute a finite-state abstraction of the dynamical system.
We obtain such an abstract model from a \emph{partition} of the continuous state space into a set of disjoint convex \emph{regions}.
Actions in this abstraction correspond to control inputs that induce transitions between these regions.
Due to the process noise, the outcome of an action is stochastic, and every transition has a certain probability.

\subsubsection{Probability intervals.}
Since the distribution of the noise is unknown, it is not possible to compute the transition probabilities exactly.
Instead, we estimate the probabilities based on a finite number of \emph{samples} (also called scenarios) of the noise, which may be obtained from a high fidelity (blackbox) simulator or from experiments.
To be \emph{robust} against estimation errors in these probabilities, we adapt tools from the \emph{scenario approach} (also called scenario optimization), which is a methodology to deal with stochastic convex optimization in a data-driven fashion~\citep{DBLP:journals/siamjo/CampiG08, garatti2019risk}.
We compute \emph{upper and lower bounds} on the transition probabilities with a desired \emph{confidence level}, which we choose up front.
These bounds are \emph{PAC}, as they contain the true probabilities with at least this confidence level.

\subsubsection{Interval MDPs.}
We formalize our abstractions with the \gls{PAC} probability bounds using so-called \glspl{iMDP}.
While regular \acrshortpl{MDP} require precise transition probabilities, \glspl{iMDP} exhibit probability intervals~\cite{DBLP:journals/ai/GivanLD00}. 
Policies for \glspl{iMDP} have to robustly account for all possible probabilities within the intervals, and one usually provides upper and lower bounds on maximal or minimal reachability probabilities or expected rewards~\citep{DBLP:conf/qest/HahnHHLT17, DBLP:conf/cav/PuggelliLSS13, DBLP:conf/cdc/WolffTM12}.
For \acrshortpl{MDP} with precise probabilities, mature tool support exists, for instance, via \prism~\citep{DBLP:conf/cav/KwiatkowskaNP11}. 
In this work, we extend the support of \prism to \glspl{iMDP}.

\subsubsection{Iterative abstraction scheme.}
The tightness of the probability intervals
depends on the number of noise samples. 
Hence, we propose an \emph{iterative abstraction scheme}, to iteratively improve these intervals by using increasing sample sizes.
For the resulting \gls{iMDP}, we compute a robust policy that maximizes the probability to safely reach the goal states.
Based on a pre-defined threshold, we decide whether this probability is unsatisfactory or satisfactory. 
In the former case, we collect additional samples to reduce the uncertainty in the probability intervals.
If the probability is satisfactory, we use the policy to compute a controller for the dynamical system.
The specified confidence level reflects the likelihood that the optimal reachability probability on the \gls{iMDP} is a \emph{lower bound} for the probability that the dynamical system satisfies the reach-avoid problem under this derived controller.

\subsubsection{Contributions.}
Our contributions are threefold:
(1)~We propose a novel method to compute safe controllers for dynamical systems with unknown noise distributions.
Specifically, the probability of safely reaching a target is guaranteed, with high confidence, to exceed a pre-defined threshold.
(2)~We propose a scalable refinement scheme that incrementally improves the \gls{iMDP} abstraction by iteratively increasing the number of samples.
(3)~We apply our method to multiple realistic control problems, and benchmark against two other tools: \stochy and \sreachtools.
We demonstrate that the guarantees obtained for the \gls{iMDP} abstraction carry over to the dynamical system of interest.
Moreover, we show that using probability intervals instead of point estimates of probabilities yields significantly more robust results.

\subsection*{Related work}

\subsubsection{Reachability analysis.}
Verification and controller synthesis for reachability in stochastic systems is an active field of research in safety-critical engineering~\citep{Abate2008probabilisticSystems,LSAZ21}.
Most approaches are based on formal abstractions~\citep{Alur2000,DBLP:journals/tac/LahijanianAB15,DBLP:journals/siamads/SoudjaniA13} or work in the continuous domain directly, e.g., using Hamilton-Jacobi reachability analysis~\citep{DBLP:conf/cdc/BansalCHT17,DBLP:conf/cdc/HerbertCHBFT17} or optimization~\citep{Rosolia2020unified}.
Several tools exist, such as \stochy~\citep{DBLP:conf/tacas/CauchiA19}, \probreach~\citep{DBLP:conf/hybrid/ShmarovZ15} and \sreachtools~\citep{DBLP:conf/hybrid/VinodGO19}.
However, the majority of these methods require full knowledge of the models.

We break away from this literature in putting forward abstractions that \emph{do not require any knowledge of the noise distribution}, via the \emph{scenario approach}.
It has been used for the verification of \glspl{MDP} with uncertain parameters~\citep{DBLP:conf/tacas/Cubuktepe0JKT20}, albeit only for finite-state systems.
\sreachtools also exhibits a sampling-based method, but relies on Hoeffding's inequality to obtain confidence guarantees~\citep{DBLP:conf/amcc/SartipizadehVAO19}, so the noise is still assumed to be sub-Gaussian~\citep{DBLP:books/daglib/0035704}.
By contrast, the scenario approach is \emph{completely distribution-free}~\citep{campi2018introduction}.
Moreover, \sreachtools is limited to problems with convex safe sets (a restrictive assumption in many problems) and its sampling-based methods can only synthesize open-loop controllers.
Further related are sampling-based feedback motion planning algorithms, such as LQR-Trees.
However, sampling in LQR-Trees relates to random exploration of the state space, and not to stochastic noise affecting the dynamics as in our setting~\citep{DBLP:journals/ijrr/ReistPT16,DBLP:conf/rss/Tedrake09}.

\subsubsection{Alternatives to the scenario approach.}
Monte Carlo methods (e.g. particle methods) can also solve stochastic reach-avoid problems~\citep{DBLP:journals/trob/BlackmoreOBW10,DBLP:conf/cdc/LesserOE13}.
These methods simulate the system via many samples of the uncertain variable~\citep{Smith2013sequentialMonteCarlo}.
Monte Carlo methods \emph{approximate} stochastic problems, while our approach provides \emph{bounds} with a desired \emph{confidence level}.

In \gls{DRO}, decisions are robust with respect to \emph{ambiguity sets} of distributions~\citep{DBLP:journals/mp/EsfahaniK18, goh2010distributionally, DBLP:journals/ior/WiesemannKS14}.
While the scenario approach uses samples of the uncertain variable, \gls{DRO} works on the domain of uncertainty directly, thus involving potentially complex ambiguity sets~\citep{garatti2019risk}.
Designing robust policies for \glspl{iMDP} with known uncertainty sets was studied by~\citeauthor{DBLP:conf/cav/PuggelliLSS13} (\citeyear{DBLP:conf/cav/PuggelliLSS13}), and~\citeauthor{DBLP:conf/cdc/WolffTM12} (\citeyear{DBLP:conf/cdc/WolffTM12}). 
Hybrid methods between the scenario approach and robust optimization also exist~\citep{Margellos2014OnProblems}.

\subsubsection{PAC literature.}
The term \gls{PAC} refers to obtaining, with high probability, a hypothesis that is a good approximation of some unknown phenomenon~\citep{DBLP:conf/aaai/Haussler90}.
\gls{PAC} learning methods for discrete-state \glspl{MDP} are developed in \citeauthor{DBLP:journals/jmlr/BrafmanT02} (\citeyear{DBLP:journals/jmlr/BrafmanT02}),~\citeauthor{DBLP:conf/rss/FuT14} (\citeyear{DBLP:conf/rss/FuT14}), and~\citeauthor{DBLP:journals/ml/KearnsS02} (\citeyear{DBLP:journals/ml/KearnsS02}), and \gls{PAC} statistical model checking for \glspl{MDP} in~\citeauthor{DBLP:conf/cav/AshokKW19} (\citeyear{DBLP:conf/cav/AshokKW19}).

\subsubsection{Safe learning methods.}
We only briefly discuss the emerging field of safe learning~\citep{Brunke2021safelearning,DBLP:journals/jmlr/GarciaF15}.
Recent works use Gaussian processes for learning-based model predictive control~\citep{DBLP:journals/tcst/HewingKZ20,DBLP:conf/cdc/KollerBT018} or reinforcement learning with safe exploration~\citep{DBLP:conf/nips/BerkenkampTS017}, and control barrier functions to reduce model uncertainty~\citep{DBLP:conf/l4dc/TaylorSYA20}.
Safe learning control concerns learning unknown, \emph{deterministic} system dynamics, while imposing strong assumptions on stochasticity~\citep{DBLP:journals/tac/FisacAZKGT19}.
By contrast, our problem setting is fundamentally different: we reason about \emph{stochastic} noise of a completely unknown distribution.
\section{Foundations and Outline}

\label{sec:Preliminaries}

A \emph{discrete probability distribution} over a finite set $X$ is a function $\mathit{prob} \colon X \to [0,1]$ with $\sum_{x \in X} \mathit{prob}(x) = 1$.
The set of all distributions over $X$ is $\distr{X}$, and the cardinality of a set $X$ is $|X|$.
A \emph{probability density function} over a random variable $x$ conditioned on $y$ is written as $p(x \vert y)$.
All vectors $\bm{x} \in \R^n$, $n \in \N$, are column vectors and denoted by bold letters.
We use the term \emph{controller} when referring to dynamical systems, while we use \emph{policy} for (i)\glspl{MDP}.

\subsection{Linear dynamical systems}
\label{subsec:LinDynSystems}

We consider discrete-time, continuous-state systems, where the progression of the $n$-dimensional state $\bm{x} \in \R^n$ depends \emph{linearly} on the current state, a control input, and a process noise term.
Given a state $\bm{x}_k$ at discrete time step $k \in \N$, the successor state at time $k+1$ is computed as
\begin{equation}
    \bm{x}_{k+1} = A \bm{x}_{k} + B \bm{u}_{k} + \bm{q}_k + \bm{w}_k,
    \label{eq:continuousSystem}
\end{equation}
where $\bm{u}_k \in \cControlSpace \subset \R^p$ is the control input at time $k$, $A \in \R^{n \times n}$ and $B \in \R^{n \times p}$ are appropriate matrices, $\bm{q}_k \in \R^n$ is a deterministic disturbance, and $\bm{w}_k \in \R^n$ is an arbitrary additive process noise term.
We consider \emph{piecewise-linear feedback controllers} of the form $\controller \colon \R^n \times \N \to \cControlSpace$, which map a state $\bm{x}_k \in \R^n$ and a time step $k \in \N$ to a control input $\bm{u}_k \in \cControlSpace$.
The controller may be time-dependent, because we consider control objectives with a finite time horizon.

The random variable $\bm{w}_k \in \Delta$ is defined on a probability space $(\Delta, \mathcal{D}, \amsmathbb{P})$, with $\sigma$-algebra $\mathcal{D}$ and probability measure $\amsmathbb{P}$ defined over $\mathcal{D}$.
We do not require the sample space $\Delta$ and probability measure $\amsmathbb{P}$ to be known explicitly. 
Instead, we employ a sampling-based approach, for which it suffices to have a finite number of $N$ \gls{iid} samples of the random variable, and to assume that its distribution is independent of time.
Due to the process noise $\bm{w}_k$, the successor state $\bm{x}_{k+1}$ is a random variable at time $k$.
We denote the probability density function over successor states as $p_{\bm{w}_k}(\bm{x}_{k+1} \ | \ \hat{\bm{x}}_{k+1})$, where $\hat{\bm{x}}_{k+1} = A \bm{x}_k + B \bm{u}_k + \bm{q}_k$ is its \emph{noiseless value}.

\begin{remark}[Restriction to linear systems]
    \label{remark:LinearSystems}
    Our methods are theoretically amenable to nonlinear systems, albeit requiring more advanced $1$-step reachability computations unrelated to our main contributions.
    Hence, we restrict ourselves to the linear system in \cref{eq:continuousSystem}.
    We discuss extensions to nonlinear systems in \cref{sec:Conclusion}.
\end{remark}

\subsubsection{Problem statement.}
We consider control objectives that are expressed as \emph{(step-bounded) reach-avoid properties}.
A reach-avoid property $\property$ is satisfied if, starting from state $\bm{x}_0$ at time $k=0$, the system reaches a desired \emph{goal region} $\goalRegion \subset \R^n$ within a finite time horizon of $\horizon \in \N$ steps, while avoiding a \emph{critical region} $\criticalRegion \subset \R^n$.
We write the probability of satisfying a reach-avoid property $\property$ under a controller $\controller$ as $\reachProbCont$.
We state the formal problem as follows.
\begin{mdframed}[backgroundcolor=gray!20, nobreak=true, innerleftmargin=8pt, innerrightmargin=8pt]
Compute a controller $\controller$ for the system in \cref{eq:continuousSystem} that, with high confidence, guarantees that $\reachProbCont \geq \eta$, where $\eta \in [0,1]$ is a pre-defined probability threshold.
\end{mdframed}

\subsection{Markov decision processes}

A \emph{\glsfirst{MDP}} is a tuple $\mdp=\MDP$ where $\States$ is a finite set of states, $\Actions$ is a finite set of actions, $\initState$ is the initial state, and $\transfunc \colon \States \times \Actions \rightharpoonup \distr{\States}$ is the (partial) probabilistic transition function.   
We call $(s,a,s')$ with probability $\transfunc(s,a)(s')>0$ a \emph{transition}.
A deterministic (or pure) policy~\citep{DBLP:books/daglib/BaierKatoen2008} for an \gls{MDP} $\mdp$ is a function $\policy \colon \States^* \to \Actions$, where $\States^*$ is a sequence of states.
The set of all possible policies for $\mdp$ is denoted by $\policySpace_\mdp$.
Note that we leave out rewards for brevity, but our approach is directly amenable to expected reward properties~\cite{DBLP:books/daglib/BaierKatoen2008}.

A \emph{probabilistic reach-avoid property} $\reachProbDiscr$ for an \gls{MDP} describes the probability of reaching a set of goal states $\goalStates \subset \States$ within $\horizon \in \N$ steps under policy $\policy \in \policySpace$, while avoiding a set of critical states $\criticalStates \subset \States$, where $\goalStates \cap \criticalStates = \varnothing$.
An optimal policy $\policy^* \in \policySpace_\mdp$ for \gls{MDP} $\mdp$ maximizes the \emph{reachability probability}:
\begin{equation}
    \policy^* = \argmax_{\policy \in \policySpace_\mdp} \reachProbDiscr.
	\label{eq:optimalPolicy}
\end{equation}
We now relax the assumption that probabilities are precisely given.
An \emph{\glsfirst{iMDP}} is a tuple $\imdp=\iMDP$ where the uncertain (partial) probabilistic transition function $\transfuncImdp \colon \States \times \Actions \times \States \rightharpoonup \Interval$ is defined over intervals $\Interval = \{ [a,b] \ | \ a,b \in (0,1] \text{ and } a \leq b \}$. 
\glspl{iMDP} define sets of \glspl{MDP} that vary only in their transition function.
In particular, for an \gls{MDP} transition function $\transfunc$, we write $\transfunc \in \transfuncImdp$ if for all $s,s' \in \States$ and $a \in \Actions$ we have $\transfunc(s,a)(s') \in \transfuncImdp (s,a)(s')$ and $P(s, a)\in\distr{\States}$.
For \glspl{iMDP}, a policy needs to be \emph{robust} against all $\transfunc \in \transfuncImdp$.
We employ value iteration to compute a policy $\policy^* \in \policySpace_{\imdp}$ for \gls{iMDP} $\imdp$ that maximizes the lower bound on the reachability probability $\reachProbDiscrMin$ within horizon $\horizon$:
\begin{equation}
	\policy^* 
	= \argmax_{\policy \in \policySpace_{\imdp}} \reachProbDiscrMin 
	= \argmax_{\policy \in \policySpace_{\imdp}} \, \min_{\transfunc \in \transfuncImdp}  \reachProbDiscr.
	\label{eq:optimalPolicy_lb}
\end{equation}
\noindent
Note that deterministic policies suffice to obtain optimal values for (i)\glspl{MDP}~\citep{DBLP:books/wi/Puterman94,DBLP:conf/cav/PuggelliLSS13}.

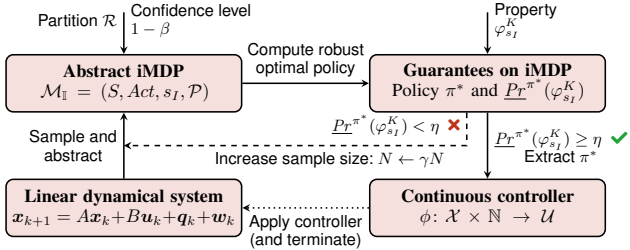
\begin{figure}[t!]
	\centering
	\usetikzlibrary{calc}
\usetikzlibrary{arrows.meta}
\usetikzlibrary{positioning}
\usetikzlibrary{fit}
\usetikzlibrary{shapes}

\tikzstyle{flowNodeRed} = [rectangle, rounded corners, minimum width=4.4cm, text width=4.3cm, minimum height=1.1cm, text centered, draw=black, fill=red!15]

\resizebox{\linewidth}{!}{%
	\begin{tikzpicture}[node distance=3.0cm,->,>=stealth,line width=0.3mm,auto,
		main node/.style={circle,draw,font=\sffamily\Large\bfseries}]
		
		\newcommand\xshift{3.8cm}
		\newcommand\xshiftB{4.0cm}
		
		\node (dynSys) [flowNodeRed] {\textbf{Linear dynamical system} \\ $\bm{x}_{k+1} = A\bm{x}_k + B\bm{u}_k + \bm{q}_k + \bm{w}_k$};
		
		\node (iMDP) [flowNodeRed, above of=dynSys, yshift=-0.6cm] {\textbf{Abstract \gls{iMDP}} \\ $\imdp = \iMDP$};
		
		\node (verify) [flowNodeRed, right of=iMDP, xshift=\xshiftB] {\textbf{Guarantees on \gls{iMDP}} \\ Policy $\policy^*$ and $\reachProbDiscrMinStar$};
		
		\node (policy) [flowNodeRed, right of=dynSys, xshift=\xshiftB] {\textbf{Continuous controller} \\ $\controller \colon \cStateSpace \times \N \to \cControlSpace$};
		
		\node (in) [above of=iMDP, yshift=-1.5cm] {};
		\node (in2) [above of=verify, yshift=-1.5cm] {};
		
		\draw [->] (in) -- (iMDP) node [pos=0.2, left, align=right, font=\sffamily\small] {Partition $\Partition$} node [pos=0.2, right, align=left, font=\sffamily\small] {Confidence level \\ $1-\beta$};
		
		\draw [->] (in2) -- (verify) node [pos=0.2, right, align=left, font=\sffamily\small] {Property\\$\propertyMDP$};
		
		\draw [->] (dynSys) -- (iMDP) node [midway, left, align=center, font=\sffamily\small] {Sample and \\ abstract};
		\draw [->] (iMDP) -- (verify) node [midway, above, align=center, font=\sffamily\small] 
		{Compute robust\\optimal policy};
		
		\draw [->, dashed] ($(verify.south) + (-0.4, 0)$) |- (0,1.2) node [pos=0.2, left, align=center, font=\sffamily\small] {$\reachProbDiscrMinStar < \eta$ \ \rured{\faIcon{times}}} node [pos=0.70, below, align=center, font=\sffamily\small] {Increase sample size: $N \gets \gamma N$};
		
		\draw [->] ($(verify.south) + (0.0, 0)$) -- ($(policy.north) + (0.0, 0)$) node [pos=0.5, right, align=center, font=\sffamily\small] {$\reachProbDiscrMinStar \geq \eta$ \ \darkgreen{\faIcon{check}} \\ Extract $\policy^*$};
		
		\draw [dotted, ->] (policy) -- (dynSys) node [midway, below, align=center, font=\sffamily\small] {Apply controller\\(and terminate)};
		
	\end{tikzpicture}
}
	
	\caption{
		Our iterative approach between abstraction and verification, where $N$ is the number of samples used for the abstraction, and $\eta$ is the threshold reachability probability.
	}
	\label{fig:Approach}
\end{figure}

\subsection{Our iterative abstraction scheme}

\label{sec:IterativeScheme}

Our proposed approach is shown in \cref{fig:Approach}.
We choose a fixed state-space partition and confidence level up front, and select an initial number of samples $N$ (note that extensions to variable confidence levels or partitions are straightforward).
As explained in \cref{sec:Abstraction,sec:ScenarioApproach}, we then abstract the dynamical system as an \gls{iMDP} using $N$ samples of the noise.
For the \gls{iMDP}, we compute an optimal policy $\policy^*$ that maximizes the probability of satisfying the given property, as per \cref{eq:optimalPolicy_lb}.
If the maximum reachability probability under this policy is above the required threshold $\eta$, we compute the corresponding controller $\controller$ for the dynamical system, and terminate the scheme.
If the maximum probability is unsatisfactory (i.e. below $\eta$), we obtain additional samples by increasing $N$ by a fixed factor $\gamma > 1$.
The updated \gls{iMDP} has tighter probability intervals, but may also have more transitions.
Since the states and actions of the \gls{iMDP} are independent of $N$, they are only computed once, in the first iteration.
In general we cannot guarantee \emph{a priori} that the property is satisfiable up to the given value of $\eta$, so we also terminate the scheme after a fixed number of iterations, in which case no output is returned.
\section{Finite-State MDP Abstraction}

\label{sec:Abstraction}

First, we describe how we partition the state space into a set of discrete convex regions.
We then use this partition to build a finite-state abstraction of the dynamical system in \cref{eq:continuousSystem}.

\subsection{State space discretization}

We choose a \emph{partition} $\Partition$ of the continuous state space $\R^n$ into a set of disjoint \emph{regions} that represent a bounded portion $\cStateSpace \subset \R^n$.
In addition, we define a single \emph{absorbing region} $r_a$, representing $\R^n \backslash \cStateSpace$.
We number the regions in $\Partition$ from $1$ to $\vert \Partition \vert$, and define a function $\Map \colon \R^n \to \{1,2,\dots,\vert \Partition \vert, \vert \Partition \vert + 1 \}$ that maps a continuous state $\bm{x} \in \R^n$ to one of the regions in partition $\Partition$ through the index of that region, or to $ \vert \Partition \vert + 1$ if $\bm{x} \in r_a = \R^n \backslash \cStateSpace$.
Thus, the absorbing region $r_a$ captures the event that the continuous state leaves the bounded portion of the state space over which we plan.
For convenience, we also define the inverse mapping as $R_i = \Map^{-1}(i)$.

We consider the regions in $\Partition$ to be $n$-dimensional bounded, convex polytopes.
In particular, convex polytope $R_i$ is the solution set of $m$ linear inequalities parameterized by $M_i \in \R^{m \times n}$ and $\bm{b}_i \in \R^m$, yielding $R_i = \big\{ \bm{x} \in \R^n \, \vert \, M_i \bm{x} \leq \bm{b}_i \big\}.$
In addition, the following assumption allows us to translate properties for the dynamical system to properties on the \gls{iMDP} abstraction:

\begin{assumption}
    \label{assumption:RegionAlignment}
    The continuous goal region $\goalRegion$ and critical region $\criticalRegion$ are aligned with the union of a subset of regions in $\Partition$, i.e. $\goalRegion = \cup_{i \in I} R_i$ and $\criticalRegion = \cup_{j \in J} R_j$ for index sets $I,J \subset \{1,2,\ldots,\card{\Partition}\}$.
\end{assumption}

\subsection{MDP abstraction}

We formalize the dynamical system discretized under partition $\Partition$ as an \gls{MDP} $\mdp = \iMDP$, by defining its states, actions, and transition probabilities (cf. ensuing paragraphs). 
We assume the initial state $\initState \in \States$ is known, and we capture time constraints by the bounded reach-avoid property.

\subsubsection{States.}
The set of states is $\States = \{ s_i \ \vert \ i = 1,\ldots,\vert\Partition\vert \} \cup \{ s_a \}$, where discrete state $s_i$ represents all continuous states $\bm{x}_k$ for which $\Map(\bm{x}_k) = i$.
Then, the \gls{MDP} consists of $\vert \States \vert = \vert \Partition \vert + 1$ states: one for every region in partition $\Partition$, plus one state $s_a$ corresponding to the absorbing region $r_a$.
State $s_a$ is a deadlock, meaning the only transition leads back to $s_a$.

\subsubsection{Actions.}
Discrete actions correspond to the execution of a control input $\bm{u}_k \in \cControlSpace$ in the dynamical system in \cref{eq:continuousSystem}.
We define $q \in \N$ \gls{MDP} actions, so $\Actions = \{ a_1, \ldots, a_q \}$.
Recall that the noiseless successor state of $\bm{x}_k$ is $\hat{\bm{x}}_{k+1} = A\bm{x}_k + B\bm{u}_k + \bm{q}_k$.
Every action $a_j$ is associated with a fixed continuous \emph{target point} $\bm{d}_j \in \R^n$, and is defined such that its noiseless successor state $\hat{\bm{x}}_{k+1} = \bm{d}_j$.
While not a restriction of our approach, we define one action for every \gls{MDP} state, and choose the target point to be the center of its region.

The \gls{MDP} must form a \emph{correct abstraction} of the dynamical system.
Thus, action $a_j$ only exists in an \gls{MDP} state $s_i$ if, for every continuous state $\bm{x}_k \in R_i$, there exists a control $\bm{u}_k \in \cControlSpace$, such that $\hat{\bm{x}}_{k+1} = \bm{d}_j$.
To impose this constraint, we define the \emph{one-step backward reachable set} $\mathcal{G}(\bm{d}_j)$:
\begin{equation}
    \mathcal{G}(\bm{d}_j)
	= \{ \bm{x} \in \R^n \,\, | 
	\,\, \bm{d}_j = A \bm{x} + B \bm{u}_k + \bm{q}_k, 
	\,\, \bm{u}_k \in \cControlSpace \}.
	\label{eq:BackwardReach}
\end{equation}
Then, action $a_j$ exists in state $s_i$ if and only if $R_i \subseteq \mathcal{G}(\bm{d}_j)$.
Note that the existence of an action in an \gls{MDP} state merely implies that \emph{for every continuous state} in the associated region, \emph{there exists} a feasible control input that induces this transition.
The following assumption asserts that the regions in $\Partition$ can indeed be contained in the backward reachable set.
\begin{assumption}
    \label{assump:BackwardReachRank}
    The backward reachable set $\mathcal{G}(\bm{d}_j)$ has a non-empty interior, which implies that matrix $B$ is full row rank, i.e., $\rank{B} = n$, where $n=\dim(\bm{x})$ in \cref{eq:continuousSystem}.
\end{assumption}
For many systems, we may group together multiple discrete time steps in \cref{eq:continuousSystem}, such that \cref{assump:BackwardReachRank} holds (see~\citeauthor{Badings2021FilterUncertainty} (\citeyear[Sect. 6]{Badings2021FilterUncertainty}) for more details).
To compute the actual control $\bm{u}_k$ in state $\bm{x}_k$ at time $k$, we replace $\bm{x}_{k+1}$ by $\bm{d}_j$ in \cref{eq:continuousSystem} and solve for $\bm{u}_k$, yielding:
\begin{equation}
    \bm{u}_k = B^+ (\bm{d}_j - \bm{q}_k - A\bm{x}_k) ,
    \label{eq:controlLaw}
\end{equation}
with $B^+$ the pseudoinverse of $B$.
It is easily verified that for every state where action $a_j$ is enabled, there exists a $\bm{u}_k$ such that \cref{eq:controlLaw} holds (depending on $B$, it may not be unique).

\subsubsection{Transition probability intervals.}
We want to determine the probability $P(s_i, a_l)(s_j)$ to transition from state $s_i$ to state $s_j$ upon choosing action $a_l$.
In the abstraction, this is equivalent to computing the \emph{cumulative density function} of the distribution over the successor state $\bm{x}_{k+1}$ under the polytope $R_j$ associated with state $s_j$.
The probability density function $p_{\bm{w}_k}(\bm{x}_{k+1} \ | \ \hat{\bm{x}}_{k+1} = \bm{d}_j )$ captures the distribution over successor states $\bm{x}_{k+1}$, which depends on the process noise $\bm{w}_k$.
By denoting $P_{\bm{w}_k}(\bm{x}_{k+1} \in R_j)$ as the probability that $\bm{x}_{k+1}$ takes a value in discrete region $R_j$, we write:
\begin{equation}
\begin{split}
    P(s_i, a_l)(s_j) &= P_{\bm{w}_k}(\bm{x}_{k+1} \in R_j) 
    \\
    &= \int_{R_j} p_{\bm{w}_k}(\bm{x}_{k+1} \ | \ \hat{\bm{x}}_{k+1} = \bm{d}_j )d\bm{x}_{k+1}.
	\label{eq:multivariateIntegral_mainProblem}
\end{split}
\end{equation} 
Recall that the probability density function $p_{\bm{w}_k}(\cdot)$ is unknown, making a direct evaluation of \cref{eq:multivariateIntegral_mainProblem} impossible.
Instead, we use a sampling-based approach to compute \textit{probability intervals} as explained in \cref{sec:ScenarioApproach}.
\section{Sampling-Based Probability Intervals}

\label{sec:ScenarioApproach}

We introduce a sampling-based method to estimate the transition probabilities in \cref{eq:multivariateIntegral_mainProblem}, based on a finite set of $N$ observations $\bm{w}_k^{(i)} \in \Delta, \ i = 1, \ldots, N$ of the process noise.
Each sample has a unique index $i = 1, \ldots, N$ and is associated with a possible successor state $\bm{x}_{k+1} = \hat{\bm{x}}_{k+1} + \bm{w}_k^{(i)}$. 
We assume that these samples are available from experimental data or simulations, and are thus obtained at a low cost.

\begin{assumption}
    \label{assump:iid}
    The noise samples $\bm{w}_k^{(i)} \in \Delta, \ i = 1, \ldots, N$ are \gls{iid} elements from $(\Delta, \amsmathbb{P})$, and are independent of time.
\end{assumption}

Due to the samples being \gls{iid}, the set $\bm{w}_k^{(1)}, \ldots, \bm{w}_k^{(N)}$ of $N$ samples is a random element from the probability space $\Delta^N$ equipped with the product probability $\amsmathbb{P}^N$.

As an example, we want to evaluate the probability $P(s_i, a_l)(s_j)$ that state-action pair $(s_i, a_l)$ induces a transition to state $s_j$.
A naive \emph{frequentist} approach to approximate the probability would be to determine the fraction of the samples leading to this transition, using the following definition.
\begin{definition}
    \label{def:Nj_in}
    The cardinality $N_j^\text{in} \in \{0,\ldots,N\}$ 
    of the index set of the samples leading to $\bm{x}_{k+1} \in R_j$ is defined as
    \begin{equation}
        N_j^\text{in} = \Big\vert \{ i \in \{1,\ldots,N\} \, \vert \, (\hat{\bm{x}}_{k+1} + \bm{w}_k^{(i)}) \in R_j \} \Big\vert. 
        \label{eq:Nj_in}
    \end{equation}
    Similarly, we define $N_j^\text{out} = N - N_j^\text{in}$ as the number of samples for which $\hat{\bm{x}}_{k+1} + \bm{w}_k^{(i)}$ is \emph{not} contained in $R_j$.
\end{definition}

Note that $N_j^\text{in}$ and $N_j^\text{out}$ depend on both the sample set and the action.
The frequentist approach is simple, but may lead to estimates that deviate critically from their true values if the number of samples is limited (we illustrate this issue in practice in \cref{subsec:UAV}).
In what follows, we introduce our method to be robust against such estimation errors.

\subsection{Bounds for the transition probabilities}

We adapt methods from the scenario approach~\citep{campi2018introduction} to compute \emph{intervals of probabilities} instead of precise estimates.
For every probability $P(s_i, a_l)(s_j)$, we compute an upper and lower bound (i.e. an interval) that contains the true probability in \cref{eq:multivariateIntegral_mainProblem} with a high confidence.
We formalize the resulting abstraction as an \gls{iMDP}, where these probability intervals enter the uncertain transition function $\transfuncImdp \colon \States \times \Actions \times \States \rightharpoonup \Interval$.
As the intervals are \gls{PAC}, this \gls{iMDP} is a robust abstraction of the dynamical system.

First, we introduce the concept of \emph{risk} (or \emph{violation probability}), which is a measure of the probability that a successor state \emph{is not} in a given region~\citep{DBLP:journals/siamjo/CampiG08}.
\begin{definition}
    \label{def:risk}
    The risk $P_{\bm{w}_k}(\bm{x}_{k+1} \notin R_j)$ that a successor state $\bm{x}_{k+1}$ is not in region $R_j$ is
    \begin{equation}
    \begin{split}
        P_{\bm{w}_k}(\bm{x}_{k+1} \notin R_j) &= \amsmathbb{P} \{ \bm{w}_k \in \Delta \ \colon \ \hat{\bm{x}}_k + \bm{w}_k \notin R_j \} 
        \\
        &= 1 - P_{\bm{w}_k}(\bm{x}_{k+1} \in R_j). 
        \label{eq:ViolationProbability}
    \end{split}
    \end{equation}
\end{definition}

Crucially for our approach, note that $P(s_i, a_l)(s_j) = P_{\bm{w}_k}(\bm{x}_{k+1} \in R_j)$.
The scenario approach enables us to bound the risk that the optimal point of a so-called \emph{scenario optimization problem} does not belong to a feasible set $\tilde{R}$ defined by a set of constraints when we are only able to sample a subset of those constraints.
By formulating this optimization problem such that $\tilde{R}$ is closely related to a region $R_j$, we obtain upper and lower bounds on the risk over $R_j$, and thus also on the corresponding transition probability 
\ifappendix
    (see \cref{appendix:Proofs} for details).
\fi
Importantly, this result means that we can adapt the theory from the scenario approach to compute transition probability intervals for our abstractions.

Based on this intuition, we state the main contribution of this section, as a non-trivial variant of~\citeauthor{DBLP:conf/cdc/RomaoMP20} (\citeyear[Theorem 5]{DBLP:conf/cdc/RomaoMP20}), adapted for our context.
Specifically, for a given transition $(s_i,a_l,s_j)$ and the corresponding number of samples $N_j^\text{out}$ outside of region $R_j$ (as per \cref{def:Nj_in}), \cref{theorem:Bounds} returns an interval that contains $P(s_i, a_l)(s_j)$ with at least a pre-defined confidence level.
\begin{theorem}[\gls{PAC} probability intervals]
    \label{theorem:Bounds}
    For $N \in \N$ samples of the noise, fix a confidence parameter $\beta \in (0,1)$.
    Given $N_j^\text{out}$, the transition probability $P(s_i, a_l)(s_j)$ is bounded by
    \begin{equation}
        \amsmathbb{P}^N \Big\{ \munderbar{p} \leq P(s_i, a_l)(s_j) \leq \bar{p} \Big\} \geq 1 - \beta,
        \label{eq:TheoremBounds}
    \end{equation}
    where $\munderbar{p} = 0$ if $N_j^\text{out} = N$, and otherwise $\munderbar{p}$ is the solution of
    \begin{equation}
        \frac{\beta}{2 N} = \sum_{i=0}^{N_j^\text{out}} \binom Ni (1-\munderbar{p})^i \munderbar{p}^{N-i},
        \label{eq:TheoremLowerBound}
    \end{equation}
    and $\bar{p} = 1$ if $N_j^\text{out} = 0$, and otherwise $\bar{p}$ is the solution of 
    \begin{equation}
        \frac{\beta}{2 N} = 1 - \sum_{i=0}^{N_j^\text{out} - 1} \binom Ni (1-\bar{p})^i \bar{p}^{N-i}.
        \label{eq:TheoremUpperBound}
    \end{equation}
\end{theorem}
\ifappendix
    For brevity, we postpone the proof and technical details to \cref{appendix:Proofs}.
\fi
\cref{theorem:Bounds} states that with a probability of at least $1-\beta$, the probability $P(s_i, a_l)(s_j)$ is bounded by the obtained interval.
Importantly, this claim holds for \emph{any} $\Delta$ and $\amsmathbb{P}$, meaning that we can bound the probability in \cref{eq:multivariateIntegral_mainProblem}, even when the probability distribution of the noise is unknown.

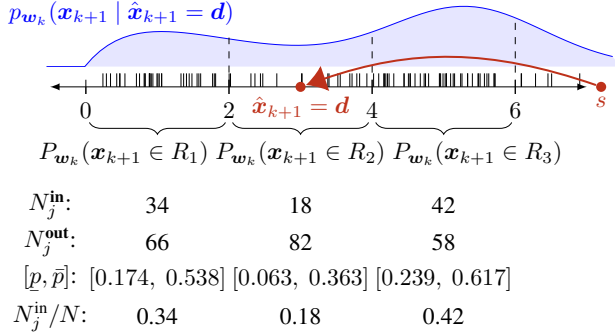
\begin{figure}[t!]
	\centering
	\resizebox{\linewidth}{!}{%

\begin{tikzpicture}[xscale=1.06, yscale=1]

\newcommand\normMean{4}
\newcommand\normScale{1}
\newcommand\gammaA{2}

\begin{axis}[
        at={(-3.55cm,0cm)},
        domain = -0.5:7,
        axis line style={draw=none},
        tick style={draw=none},
        ylabel = {},
        ylabel style={align=center,rotate=-90},
        xticklabels={,,},
        yticklabels={,,},
        yshift=0.3cm,
        xmin=-0.5, xmax=7,
        ymin=0,  ymax=0.26,
        width=9.6cm,
        height=2.5cm,
    ]
    \addplot [name path=A, color=blue] table [x=x, y=y, col sep=comma, mark=none] {Figures/Tikz/distributionPlot_sparse3.csv};
    
    \path[name path=B] (axis cs:-0.5,0) -- (axis cs:7,0);
    
    \addplot [
        thick,
        color=blue,
        fill=blue, 
        fill opacity=0.1
    ] fill between[of=A and B, soft clip={domain=-1:7},];
    
\end{axis}

\foreach \x in {1.9904310603088948,-0.5278880533336503,-1.4461112880916378,-1.5943995469103731,2.6294506104614133,-1.1867788845432967,2.037181532238858,-1.3959068973420097,2.5688452626954277,3.901003799214978,-1.5563206532574538,1.091929144540357,-1.2292227993597358,-2.063392996771781,2.255592948953187,0.6881758404254068,-2.5687775314312766,-2.2498452065637062,2.0199693617488226,1.956825926204485,0.7860453702328734,-2.52772142325479,-2.1858132538399264,2.334055970212992,2.4005500055531614,0.5210108316624362,-2.655417619739051,2.706090565629543,0.8253038068962786,-0.3318863501018159,1.7266469820105135,-1.9415769023599694,-1.5177303444790482,-2.522985295885876,-1.9401966127636552,2.479133323201429,1.6728477125759245,-0.978864651407358,0.5541711976103079,1.0204948216450314,3.5081135776747,1.4445949955555575,-2.4539701417083117,2.132481012567254,1.1618219654722655,1.5341968606117327,1.5991518762340222,-0.6935627554834376,1.3025740111999102,2.037788529815593,1.7458974425571778,2.703316071729569,-2.7131784058516173,-2.1143043485120243,2.6037162284961353,-1.974566829520481,-0.5605911771474217,2.4874738006008466,3.2832169888506773,-1.6707645617572062,1.2142909742497245,1.1912412115649103,0.013967985175968423,2.096534415839007,2.719099849235044,2.1520972561880516,-2.1091728382097736,0.9264630029369343,3.4566694045208743,1.5286715982648778,1.946569855780397,3.087362028180986,-2.1699550902726727,1.2367856267370954,2.3343339772806964,-2.2934955905575745,-1.9932353533238478,1.9395806905152062,-1.633890382078726,-2.761491108014665,1.7890688465551463,0.3807806785316634,1.447596198644714,-2.0064272349669405,-1.2056352536204722,2.236418708772468,-2.1021263761556046,2.5086611768937415,-0.637435936712774,3.0825141682213797,-2.176049547871288,-2.110686716589787,-2.0858006381367344,0.719254321697882,-1.4211937737964444,0.021350708152708897,1.5067085167756558,-0.9983214262961626,0.3976297303859848,1.173806561264576}
{
    \draw[line width=0.01mm] (\x,0) -- ++(0,0.2cm);
}

\draw[dashed] (-1cm,0cm) -- ++(0cm,0.73cm);
\draw[dashed] (1cm,0cm) -- ++(0cm,0.80cm);
\draw[dashed] (3cm,0cm)  -- ++(0cm,1.00cm);

\draw[latex-latex] (-3.5,0) -- (4,0) ; 

\foreach \x in  {-3,-1,1,3} 
    \draw[shift={(\x,0)},color=black] (0pt,3pt) -- (0pt,-3pt);

\foreach \x/\xtext in {-3/0,-1/2,1/4,3/6} 
    \draw[shift={(\x,0)},color=black] (0pt,0pt) -- (0pt,-3pt) node[below] 
    {$\xtext$};
    
\draw (-2.5cm,1.1cm) node[align=center, blue] {$p_{\bm{w}_k}(\bm{x}_{k+1} \ \vert \ \hat{\bm{x}}_{k+1} = \bm{d})$};

\node[red, label={[red, below, label distance=-0.1cm]$s$}] (predecessor) at (4.2,0) [circle,fill,inner sep=1.5pt] {};
\node[red, label={[red, below, label distance=-0.1cm]$\hat{\bm{x}}_{k+1} = \bm{d}$}] (target) at (0,0) [circle,fill,inner sep=1.5pt] {};

\draw[-{Latex[length=3mm,width=4mm]}, line width=0.3mm, red] (predecessor)  to [out=160,in=20] (target);

\draw [decorate,decoration={brace,amplitude=5pt,mirror,raise=2ex}]
  (-2.95,-6pt) -- (-1.05,-6pt) node[midway, yshift=-2.2em, xshift=-.5cm]{$P_{\bm{w}_k}(\bm{x}_{k+1} \in R_1)$};
  
\draw [decorate,decoration={brace,amplitude=5pt,mirror,raise=2ex}]
  (-0.95,-6pt) -- (0.95,-6pt) node[midway,yshift=-2.2em]{$P_{\bm{w}_k}(\bm{x}_{k+1} \in R_2)$};
  
\draw [decorate,decoration={brace,amplitude=5pt,mirror,raise=2ex}]
  (1.05,-6pt) -- (2.95,-6pt) node[midway,yshift=-2.2em, xshift=.5cm]{$P_{\bm{w}_k}(\bm{x}_{k+1} \in R_3)$};

\newcommand\yshift{-1.75}
\newcommand\yshiftB{-2.30}
\newcommand\yshiftC{-2.85}
\newcommand\yshiftD{-3.40}
  
\draw (-3.5, \yshift) node[align=left] {$N_j^\textbf{in}$:};
\draw (-2, \yshift) node[] {34};
\draw (0,  \yshift) node[] {18};
\draw (2,  \yshift) node[] {42};

\draw (-3.5, \yshiftB) node[align=left] {$N_j^\textbf{out}$:};
\draw (-2, \yshiftB) node[] {66};
\draw (0,  \yshiftB) node[] {82};
\draw (2,  \yshiftB) node[] {58};

\draw (-3.5, \yshiftC) node[align=left] {$[\munderbar{p}, \bar{p}]$:};
\draw (-2, \yshiftC) node[] {$[0.174, \ 0.538]$};
\draw (0, \yshiftC) node[]  {$[0.063, \ 0.363]$};
\draw (2, \yshiftC) node[]  {$[0.239, \ 0.617]$};

\draw (-3.5, \yshiftD) node[align=left] {$N_j^\text{in}/N$:};
\draw (-2, \yshiftD) node[] {0.34};
\draw (0, \yshiftD) node[]  {0.18};
\draw (2, \yshiftD) node[]  {0.42};

\end{tikzpicture}

}
	
	\caption{
	    Bounds $[\munderbar{p},\bar{p}]$ on the probabilities $P(s,a)(s_j)$ for 3 regions $j \in \{ 1,2,3 \}$, using $N=100$ samples (black ticks) and $\beta=0.01$. 
	    The distribution over successor states is $p_{\bm{w}_k}(\cdot)$.
	    Point estimate probabilities are computed as $N_j^\text{in}/N$.
	}
	\label{fig:Scenario1D_b}
\end{figure}

\subsection{Practical use of \cref{theorem:Bounds}}

\label{subsec:PracticalUse}

We describe how we apply \cref{theorem:Bounds} to compute probability intervals of the \glspl{iMDP}.
For every state-action pair $(s_i, a_l)$, we obtain $N$ samples of the noise, and we determine $N_j^\text{out}$ for every $j \in \{ 1,\ldots,\vert\Partition\vert, \vert\Partition\vert+1 \}$.
Then, we invoke \cref{theorem:Bounds} for every possible successor state $s_j \in \States$, to compute the bounds on $P(s_i, a_l)(s_j)$.
\cref{fig:Scenario1D_b} shows this process, where every tick is a successor state $\bm{x}_{k+1}$ under a sample of the noise.
This figure also shows point estimates of the probabilities, derived using the frequentist approach.
If no samples are observed in a region, we assume that $P(s_i, a_l)(s_j) = 0$.

Interestingly, our method is in practice almost as simple as the frequentist approach, but has the notable advantage that we obtain robust intervals of probabilities.
Note that \cref{eq:TheoremLowerBound,eq:TheoremUpperBound} are cumulative distribution functions of a beta distribution with parameters $N_j^\text{out} + 1$ (or $N_j^\text{out}$) and $N - N_j^\text{out}$ (or $N - N_j^\text{out} - 1$), respectively \cite{campi2018introduction}, which can directly be solved numerically for $\munderbar{p}$ (or $\bar{p}$).
To speed up the computations at run-time, we apply a tabular approach to compute the intervals for all relevant values of $N$, $\beta$, and $k$ up front.
\ifappendix
    In \cref{example:ScenarioApproach}, we exemplify how the number of samples controls the tightness of the intervals.
\fi
\begin{algorithm}[t!]
\caption{Sampling-based iMDP abstraction.}
\label{alg:algorithm}
\textbf{Input}: Linear dynamical system; property $\propertyMDP$ (threshold $\eta$)\\
\textbf{Params}: Partition $\Partition$; confidence lvl. $\beta$; increment factor $\gamma$\\
\textbf{Output}: Controller $\controller$
\begin{algorithmic}[1] 
\STATE Define iMDP states $\States$ and set of enabled actions $\Actions$
\STATE Let initial number of samples $N = N_0$, 
\STATE Let iteration limit $z^\text{max}$ and $z = 0$
\WHILE{$\reachProbDiscrMinStar < \eta$ \textbf{and} $z < z^\text{max}$}
    \FORALL{actions $a$ in $\Actions$}
        \FORALL{successor states $s$ in $\States$}
            \STATE Compute PAC interval on probability $P(\cdot, a)(s)$
        \ENDFOR
    \ENDFOR
    \STATE Generate \gls{iMDP} $\imdp=\iMDP$ for $N$ samples
    \STATE Compute $\policy^*$ and $\reachProbDiscrMinStar$ on $\imdp$ using \prism
    \STATE Let $N = \gamma N$
\ENDWHILE
\STATE \textbf{return} piece-wise linear controller $\controller$ based on $\policy^*$
\end{algorithmic}
\end{algorithm}

\section{Numerical Examples}

\label{sec:Studies}

We implement our iterative abstraction method in Python, and tailor the model checker \prism~\citep{DBLP:conf/cav/KwiatkowskaNP11} for \glspl{iMDP} to compute robust optimal policies.
We present a pseudocode of our method in \cref{alg:algorithm}.
At every iteration, the obtained \gls{iMDP} is fed to \prism, which computes the optimal policy associated with the maximum reachability probability, as per \cref{eq:optimalPolicy_lb}.
Our codes are available via \color{Sepia}\url{https://gitlab.science.ru.nl/tbadings/sample-abstract}\color{black}, and all experiments are run on a computer with 32 3.7GHz cores and 64 GB of RAM.
We report the performance of our method on: (1) a \gls{UAV} motion control, (2) a building temperature regulation, and (3) a spacecraft rendezvous problem.
In all benchmarks, we use \cref{theorem:Bounds} with $\beta=0.01$, and apply the iterative scheme with $\gamma = 2$, starting at $N = 25$, with an upper bound of $12,800$ samples.

\begin{figure}[t!] 
    \centering
    \includegraphics[width=\linewidth]
    {Figures/Results/UAV/UAV_v3.png}
    \caption{UAV problem (goal in green; obstacles in red), plus trajectories under the optimal \gls{iMDP}-based controller from $\bm{x}_0 = [-14,0,6,0,-6,0]^\top$, under high and low turbulence.}
    \label{fig:UAV_layout}
\end{figure}

\subsection{UAV motion planning}
\label{subsec:UAV}

We consider the reach-avoid problem for a \gls{UAV} operating under turbulence, which was introduced in \cref{sec:Introduction}.
Our goal is to compute a controller that guarantees (with high confidence) that the probability to reach a goal area, while also avoiding unsafe regions, is above a performance threshold of $\eta = 0.75$.
We consider a horizon of $64$ time steps, and the problem layout is displayed in \cref{fig:UAV_layout}, with goal and critical regions shown in green and red, respectively.
We model the \gls{UAV} as a system of 3 double integrators
\ifappendix
    (see \cref{appendix:UAV} for details).
\fi
The state $\bm{x}_k \in \R^6$ encodes the position and velocity components, and control inputs $\bm{u}_k \in \R^3$ model actuators that change the velocity.
The effect of turbulence on the state causes (non-Gaussian) process noise, which we model using a Dryden gust model~\citep{bohn2019deep, dryden1943review}.
We compare two cases: a) a low turbulence case, and 2) a high turbulence case.
We partition the state space into $25,515$ regions.

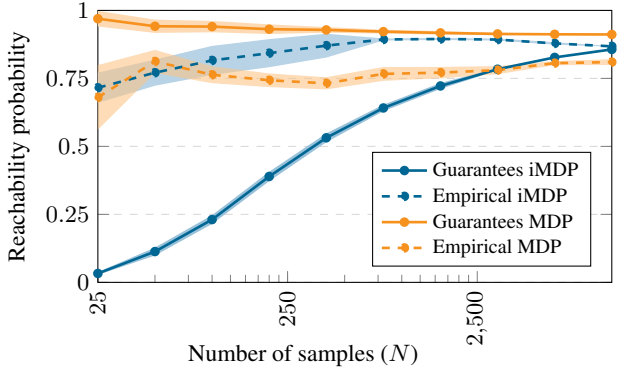
\begin{figure}[t!]
    \centering
        \footnotesize\begin{tikzpicture}
      \begin{axis}[
          width=\linewidth,
          height=5.2cm,
          ymajorgrids,
          grid style={dashed,gray!30},
          xlabel=Number of samples ($N$),
          ylabel=Reachability probability,
          x label style={at={(axis description cs:0.4,-0.2)}},
          x tick label style={rotate=90,anchor=east},
          xmode=log,
          log ticks with fixed point,
          xmin=25,
          xmax=12800,
          ymin=0,
          ymax=1,
          xtick={25,250,2500},
          xtick pos=left,
          ytick={0,0.25,...,1.0},
          every axis plot/.append style={line width=1pt},
          legend cell align={left},
          legend columns=1,
          legend style={at={(0.96,0.48)},
                        nodes={scale=0.85, transform shape},
                        anchor=north east, 
                        column sep=0ex,}
        ]
        
        \addplot[mark=*, mark size=1.3pt, color=MidnightBlue] table[x=x, y=imdp, col sep=comma] {Figures/Results/UAV/MDP_vs_iMDP_v3.csv};
        
        \addplot [name path=upper_imdp, draw=none] table[x=x, y=imdp_error_plus, col sep=comma, forget plot] {Figures/Results/UAV/MDP_vs_iMDP_v3.csv};
        \addplot [name path=lower_imdp, draw=none] table[x=x, y=imdp_error_min, col sep=comma, forget plot] {Figures/Results/UAV/MDP_vs_iMDP_v3.csv};
        \addplot [fill=MidnightBlue!30, forget plot] fill between[of=upper_imdp and lower_imdp];
        
        \addplot[mark=*, mark size=1.3pt, color=MidnightBlue, dashed] table[x=x, y=imdp_sim, col sep=comma] {Figures/Results/UAV/MDP_vs_iMDP_v3.csv};
        
        \addplot [name path=upper_imdp_sim, draw=none] table[x=x, y=imdp_sim_error_plus, col sep=comma, forget plot] {Figures/Results/UAV/MDP_vs_iMDP_v3.csv};
        \addplot [name path=lower_imdp_sim, draw=none] table[x=x, y=imdp_sim_error_min, col sep=comma, forget plot] {Figures/Results/UAV/MDP_vs_iMDP_v3.csv};
        \addplot [fill=MidnightBlue!20, forget plot, fill opacity=1.0] fill between[of=upper_imdp_sim and lower_imdp_sim];
        
        
         \addplot[mark=*, mark size=1.3pt, color=BurntOrange] table[x=x, y=mdp, col sep=comma] {Figures/Results/UAV/MDP_vs_iMDP_v3.csv};
        
        \addplot [name path=upper_mdp, draw=none] table[x=x, y=mdp_error_plus, col sep=comma, forget plot] {Figures/Results/UAV/MDP_vs_iMDP_v3.csv};
        \addplot [name path=lower_mdp, draw=none] table[x=x, y=mdp_error_min, col sep=comma, forget plot] {Figures/Results/UAV/MDP_vs_iMDP_v3.csv};
        \addplot [fill=BurntOrange!30, forget plot] fill between[of=upper_mdp and lower_mdp];
        
        \addplot[mark=*, mark size=1.3pt, color=BurntOrange, dashed] table[x=x, y=mdp_sim, col sep=comma] {Figures/Results/UAV/MDP_vs_iMDP_v3.csv};
        
        \addplot [name path=upper_mdp_sim, draw=none] table[x=x, y=mdp_sim_error_plus, col sep=comma, forget plot] {Figures/Results/UAV/MDP_vs_iMDP_v3.csv};
        \addplot [name path=lower_mdp_sim, draw=none] table[x=x, y=mdp_sim_error_min, col sep=comma, forget plot] {Figures/Results/UAV/MDP_vs_iMDP_v3.csv};
        \addplot [fill=BurntOrange!50, forget plot, fill opacity=0.6] fill between[of=upper_mdp_sim and lower_mdp_sim];

        \legend{{Guarantees iMDP},{Empirical iMDP}, {Guarantees MDP}, {Empirical MDP }}
      \end{axis}
    \end{tikzpicture}
    \caption{Reachability guarantees on the \glspl{iMDP} (blue) and \glspl{MDP} (orange) for their respective policies, versus the resulting empirical (simulated) performance (dashed lines) on the dynamical system. 
    Shaded areas show the standard deviation across 10 iterations.
    The empirical performance of the \glspl{MDP} violates the guarantees; that of the \glspl{iMDP} does not.}    
    \label{fig:MDP_vs_iMDP}
\end{figure}

\subsubsection{Scalability.}
\ifappendix
    We report the model sizes and run times in \cref{subappendix:UAV_results}.
\fi
The number of \gls{iMDP} states equals the size of the partition.
Depending on the number of samples $N$, the \gls{iMDP} has $9-24$ million transitions.
The mean time to compute the set of \gls{iMDP} actions (which is only done in the first iteration) is $\SI{15}{\minute}$.
Computing the probabilities plus the verification in \prism takes $1-\SI{8}{\minute}$, depending on the number of samples $N$.

\subsubsection{Accounting for noise matters.}
In \cref{fig:UAV_layout}, we show state trajectories under the optimal \gls{iMDP}-based controller, under high and low turbulence (noise).
Under low noise, the controller prefers the short but narrow path; under high noise, the longer but safer path is preferred.
Thus, accounting for process noise is important to obtain controllers that are safe.

\subsubsection{iMDPs yield safer guarantees than MDPs.} 
To show the importance of using robust abstractions, we compare, under high turbulence, our robust \gls{iMDP} approach against a naive \gls{MDP} abstraction.
This \gls{MDP} has the same states and actions as the \gls{iMDP}, but uses precise (frequentist) probabilities.
The maximum reachability probabilities (guarantees) for both methods are shown in \cref{fig:MDP_vs_iMDP}.
For every value of $N$, we apply the resulting controllers to the dynamical system in Monte Carlo simulations with $10,000$ iterations, to determine the empirical reachability probability.
\cref{fig:MDP_vs_iMDP} shows that the non-robust \glspl{MDP} yield \emph{poor and unsafe performance guarantees}: the actual reachability of the controller is much lower than the reachability guarantees obtained from \prism.
By contrast, our robust \gls{iMDP}-based approach consistently yields safe lower bound guarantees on the actual performance of controllers.
The performance threshold of $\reachProbDiscrMinStar \geq 0.75$ is guaranteed for $N = 3,200$ samples and higher.

\subsection{Building temperature regulation}
Inspired by \citeauthor{DBLP:journals/corr/abs-1803-06315} (\citeyear{DBLP:journals/corr/abs-1803-06315}), we consider a temperature control problem for a building with two rooms, both having their own radiator and air supply.
Our goal is to maximize the probability to reach a temperature of $\SI{20}{\celsius}$ in both zones within 32 steps of 15 minutes.
The state $\bm{x}_k \in \R^4$ of the system 
\ifappendix
    (see \cref{appendix:BAS} for details)
\fi
reflects the temperatures of both zones and radiators, and control inputs $\bm{u}_k \in \R^4$ change the air supply and boiler temperatures in both zones.
The deterministic heat gain through zone walls is modeled by the disturbance $\bm{q}_k \in \R^4$.
The noise $\bm{w}_k \in \R^4$ has a Gaussian distribution (but this assumption is not required for our approach).
We partition the state space into $35,721$ regions: $21$ values for zone temperatures and $9$ for radiator temperatures.

\begin{figure}[t!]
    \centering
    \includegraphics[width=\linewidth]{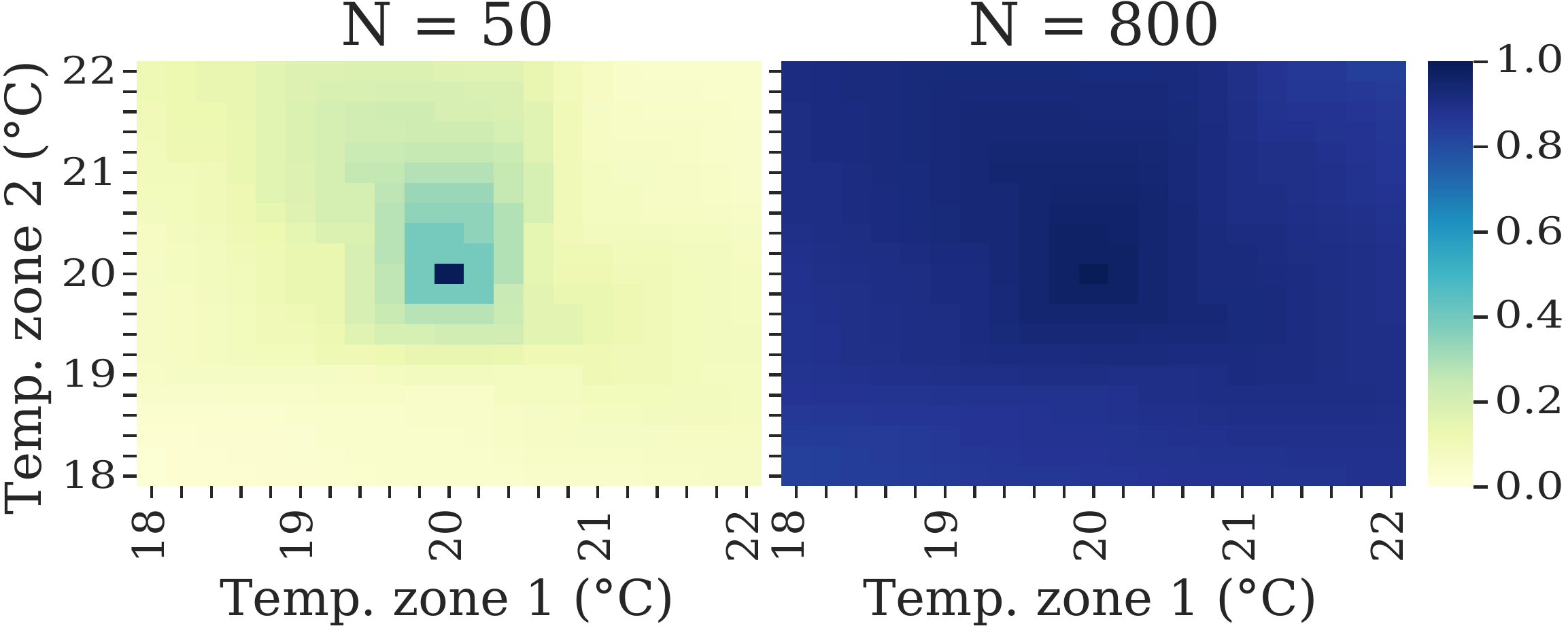}
    \caption{Cross section (for radiator temp. of \SI{38}{\celsius}) of the maximum lower bound probabilities to reach the goal of \SI{20}{\celsius} from any initial state, for either $50$ or $800$ samples.}
    \label{fig:BAS_results}
\end{figure}

\subsubsection{More samples means less uncertainty.}
In \cref{fig:BAS_results}, we show (for fixed radiator temperatures) the maximum lower bound probabilities obtained from \prism, to reach the goal from any initial state.
The results clearly show that better reachability guarantees are obtained when more samples are used to compute the \gls{iMDP} probability intervals.
The higher the value of $N$, the lower the uncertainty in the intervals, leading to better reachability guarantees.
Notably, the largest \gls{iMDP} has around $200$ million transitions, 
\ifappendix
    as reported in \cref{subappendix:BAS_results}.
\fi

\subsection{Benchmarks to other control synthesis tools}

\begin{figure}[t!]
    \centering
    \includegraphics[width=\linewidth]{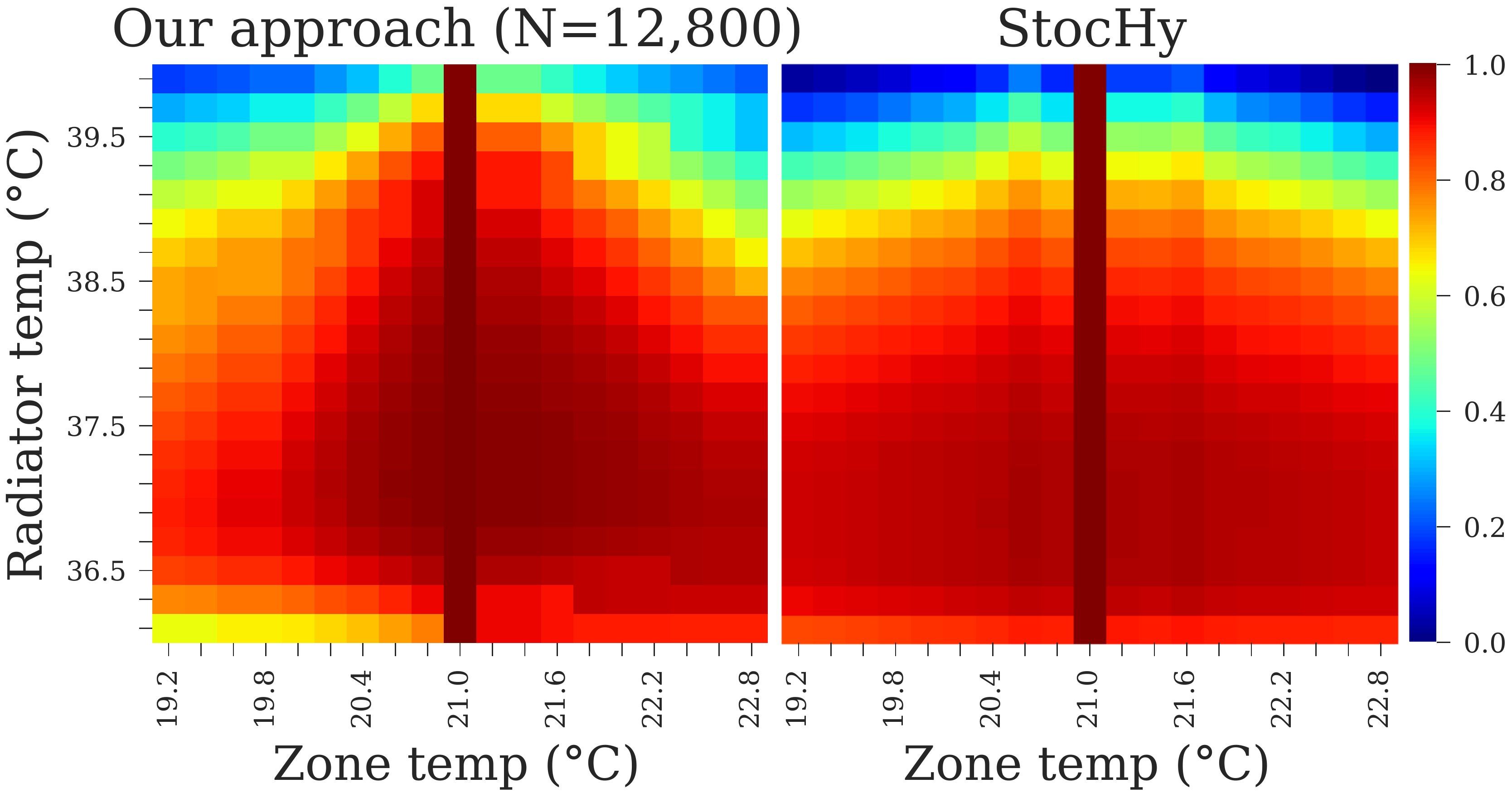}
    \caption{Maximum lower bound probabilities to reach the goal zone temperature of $\SI{21}{\celsius}$ from any initial state within $64$ steps, for our approach ($N=12,800$) and \stochy.}
    \label{fig:BAS_stochy}
\end{figure}

\subsubsection{StocHy.} 
We benchmark our method on a building temperature problem against \stochy~\citep{DBLP:conf/tacas/CauchiA19}, a verification and synthesis tool based on formal abstractions 
\ifappendix
    (see \cref{appendix:StocHy} for details on the setup and results).
\fi
Similar to our approach, \stochy also derives robust \gls{iMDP} abstractions.
However, \stochy requires precise knowledge of the noise distribution, and it discretizes the control input space of the dynamical system, to obtain a finite action space.
The maximum probabilities to reach the goal zone temperature from any initial state obtained for both methods are presented in \cref{fig:BAS_stochy}.
The obtained results are qualitatively similar, and close to the goal temperature, our lower bound reachability guarantees are \emph{slightly higher} than those obtained from \stochy.
However, when starting at temperatures close to the boundary (e.g. at both low radiator and zone temperature), the guarantees obtained from our approach are \emph{slightly more conservative}.
This is due to the fact that our approach relies on \gls{PAC} guarantees on the transition probabilities, while \stochy gives straight probabilistic outcomes.
While both methods yield results that are qualitatively similar, our approach is an order of magnitude faster ($\SI{45}{\minute}$ for \stochy, vs. $3-\SI{9}{\second}$ for our approach; 
\ifappendix
    see \cref{appendix:StocHy,tab:DetailedResults} for details).
\fi

\begin{figure}[t!]
    \centering
    \includegraphics[width=\linewidth]{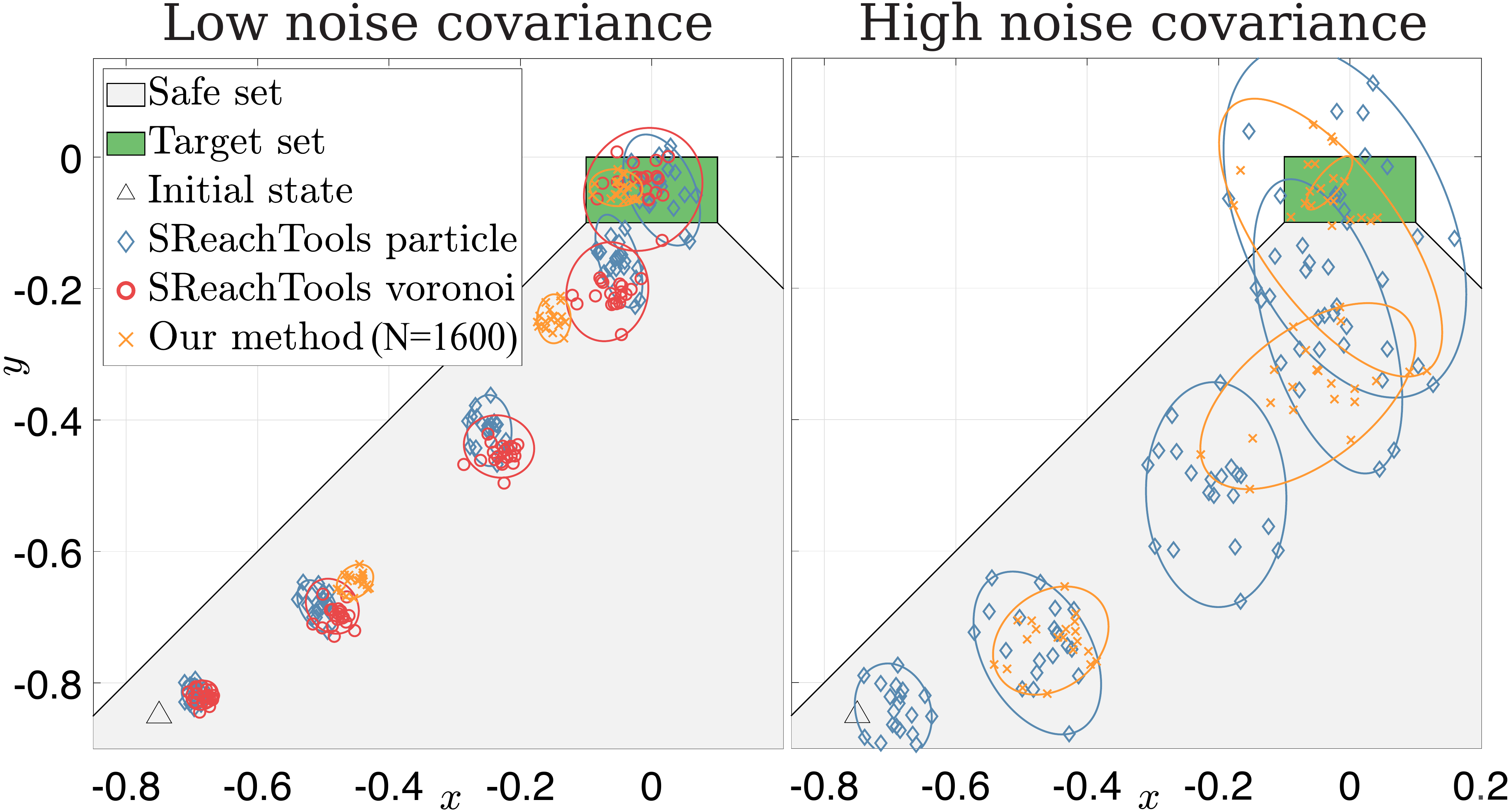}
    \caption{Simulated state trajectories for the spacecraft rendezvous problem, under low and high noise covariance.
    Our feedback controllers are more robust, as shown by the smaller error in the state trajectories over time (the Voronoi method under high covariance failed to generate a solution).}
    \label{fig:SReachTools_results}
\end{figure}

\subsubsection{\sreachtools.}
We apply our method to the spacecraft rendezvous benchmark (see \cref{fig:SReachTools_results}) of \sreachtools~\citep{DBLP:conf/hybrid/VinodGO19}, an optimization-based toolbox for probabilistic reachability problems 
\ifappendix
    (see \cref{appendix:SReachTools} for details).
\fi
While we use samples to generate a model abstraction, \sreachtools employs sample-based methods over the properties directly.
Distinctively, \sreachtools does not create abstractions (as in our case) and is thus generally faster than our method.
However, its complexity is exponential in the number of samples (versus linear complexity for our method).
Importantly, we derive \emph{feedback} controllers, while the sampling-based methods of \sreachtools compute \emph{open-loop} controllers.
Feedback controllers respond to state observations over time and are, therefore, more robust against strong disturbances from noise, as also shown in \cref{fig:SReachTools_results}.
\section{Concluding Remarks and Future Work}

\label{sec:Conclusion}

We have presented a novel sampling-based method for robust control of autonomous systems with process noise of unknown distribution.
Based on a finite-state abstraction, we have shown how to compute controllers with \gls{PAC} guarantees on the performance on the continuous system.
Our experiments have shown that our method effectively solves realistic problems and provides safe lower bound guarantees on the performance of controllers.

\subsubsection{Nonlinear systems.}
While we have focused on linear dynamical systems, as discussed in \cref{remark:LinearSystems}, we wish to develop extensions to nonlinear systems.  
Such extensions are non-trivial and may require more involved reachability computations~\citep{DBLP:conf/cdc/BansalCHT17, DBLP:conf/cav/ChenAS13}.
Specifically, the main challenge is to compute the enabled \gls{iMDP} actions via the backward reachable set defined in \cref{eq:BackwardReach}, which may become non-convex under nonlinear dynamics.
Note that computing the \gls{PAC} probability intervals remains unchanged, as the scenario approach relies on the convexity of the target set only, and not on that of the backward reachable set.
Alternatively, we may apply our method on a linearized version of the nonlinear system.
However, in order to preserve guarantees, one must then account for any linearization error.

\subsubsection{State space discretization.}
The discretization of the state space influences the quality of the reachability guarantees: a more fine-grained partition yields an abstraction that is a more accurate representation of the dynamical system, but also increases the computational complexity.
In the future, we plan to employ adaptive discretization schemes to automatically balance this trade-off, such as in~\citeauthor{DBLP:journals/siamads/SoudjaniA13} (\citeyear{DBLP:journals/siamads/SoudjaniA13}).

\subsubsection{Safe exploration.}
Finally, we wish to incorporate other uncertainties in \cref{eq:continuousSystem}, such as state/control-dependent process noise, or measurement noise.
Moreover, we may drop the assumption that the system matrices are precisely known, such that we must simultaneously learn about the unknown deterministic dynamics and the stochastic noise.
Learning deterministic dynamics is common in safe learning control~\citep{Brunke2021safelearning}, but enabling safe exploration requires strong assumptions on stochastic uncertainty.
This is a challenging goal, as it conflicts with our assumption that the distribution of the process noise is completely unknown.

\section*{Acknowledgments}
We would like to thank Licio Romao for his helpful discussions related to the scenario approach and our main theorem.

\bibliography{references, scenarioReferences}

\newpage

\ifappendix

    \appendix
    \newcommand\scalemath[2]{\scalebox{#1}{\mbox{\ensuremath{\displaystyle #2}}}}

\section{Proofs and Examples}

\label{appendix:Proofs}

In this appendix, we provide the proof of \cref{theorem:Bounds}, and we exemplify how the number of noise samples affects the tightness of the obtained intervals.
Before stating the actual proof, we introduce a number of technical details.

\subsubsection{Scaled polytopes.}
First, let us define $R_j( \lambda )$ as a version of polytope $R_j$ which is \emph{scaled} by a factor $\lambda \geq 0$ relative to a so-called Chebyshev center $\tilde{\bm{h}}_j \in \R^n$~\citep{DBLP:books/cu/BV2014}.
The resulting scaled polytope is
\begin{equation}
    R_j( \lambda ) = \{ \bm{x} \in \R^n \ \vert \ M_j\bm{x} \leq \lambda (\bm{b}_j + \tilde{\bm{h}}_j ) - \tilde{\bm{h}}_j \}.
    \label{eq:PolytopeShifted}
\end{equation}
Note that $R_j(1) = R_j$, and that shifting by $\tilde{\bm{h}}_j$ ensures that $R_j(\lambda_1) \subset R_j(\lambda_2)$ for every $0 \leq \lambda_1 < \lambda_2$.
A visualization of the scaling for an arbitrary polytope $R_j$ is shown in \cref{fig:polytopeScaling} (in this example, the Chebyshev center is not unique, since the circle can be shifted while remaining within $R_j$).

\begin{figure}[t!]
	\centering
	\usetikzlibrary{calc}
\usetikzlibrary{arrows.meta}
\usetikzlibrary{positioning}
\usetikzlibrary{fit}
\usetikzlibrary{shapes}

\tikzstyle{flowNodeRed} = [rectangle, rounded corners, minimum width=4.4cm, text width=4.3cm, minimum height=1.1cm, text centered, draw=black, fill=red!15]

\resizebox{.62\linewidth}{!}{%
	\begin{tikzpicture}[scale=2.2]
		
		\newcommand\orX{0}
        \newcommand\orY{0.3}
        \newcommand\boxW{1}
    
        \newcommand\uppX{5}
        \newcommand\uppY{3}
        
        \newcommand\originX{1}
        \newcommand\originY{1}
        \newcommand\targetX{4}
        \newcommand\targetY{2}
    	
		\newcommand\dX{\targetX+0.5*\boxW}
        \newcommand\dY{\targetY+0.5*\boxW}
    
    	\draw[red, fill=red!05] (-5/9, 7/9) -- node[anchor=north, pos=0.3] {$R_j$} (1/3, 1/3) -- (5/9, -7/9) -- (-1/3, -1/3) -- (-5/9, 7/9);
    	
    	\draw[red, fill=none, dashed] (-5/9*1.2, 7/9*1.2) -- node[anchor=south, pos=0.7, xshift=0.3cm] {$R_j(1.2)$} (1/3*1.2, 1/3*1.2) -- (5/9*1.2, -7/9*1.2) -- (-1/3*1.2, -1/3*1.2) -- (-5/9*1.2, 7/9*1.2);
    	
    	\filldraw[fill=none, blue] (0,0) circle (0.3922) node[anchor=west]{};
    	\filldraw[blue] (0,0) circle (0.02) node[black, anchor=west]{$\tilde{\bm{h}}_j$};
    	
        \draw[-{Latex[length=2mm,width=2mm]}] (-.9, -1) -- (.9, -1) node[pos=0.5, below] {$\bm{x}^{(1)}$};
        \draw[-{Latex[length=2mm,width=2mm]}] (-.9, -1) -- (-.9, 1) node[pos=0.5, left] {$\bm{x}^{(2)}$};
		
	\end{tikzpicture}
}
	
	\caption{
	    Polytope $R_j$ has a Chebyshev center $\tilde{\bm{h}}_j$ (note that it is not unique, since the circle can be shifted while remaining within the polytope).
	    Polytope $R_j(1.2)$ is scaled by a factor $\lambda = 1.2$ and is computed using \cref{eq:PolytopeShifted}.
	}
	\label{fig:polytopeScaling}
\end{figure}
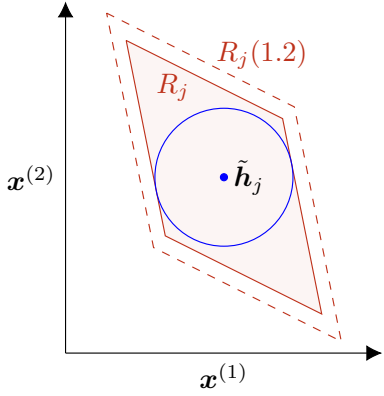

\subsubsection{Scenario optimization problem.}
We use the scenario approach to estimate the risk associated with the solution of the following convex \emph{scenario optimization problem with discarded samples}, which has a scalar decision variable $\lambda$:
\begin{align}
    \minimize_{\lambda \in \R_+} \,\, & \lambda
    \label{eq:SCP}
	\\
	\nonumber
	\text{subject to} \enskip & \hat{\bm{x}}_{k+1} + \bm{w}_k^{(i)} \in R_j (\lambda) \ \ \forall i \in \{ 1,\ldots,N \} \backslash Q,
\end{align}
whose optimal solution $\lambda_{\vert Q \vert}^*$ parameterizes polytope $R_j(\lambda_{\vert Q \vert}^*)$ through \cref{eq:PolytopeShifted}.
We explicitly write the dependency of the solution on the cardinality ${\vert Q \vert}$ of the index set $Q$, which accounts for a subset of samples whose constraints have been discarded based on the following rule:

\begin{lemma}
    The sample removal set $Q \subseteq \{ 1, \ldots, N \}$ is obtained by iteratively removing the active constraints from \cref{eq:SCP}.
    Thus, given $N$ samples and any two removal sets with cardinalities $\vert Q_1 \vert < \vert Q_2 \vert$, it holds that $Q_1 \subset Q_2$.
    Moreover, any discarded sample $i \in Q$ violates the solution $\lambda_Q^*$ to \cref{eq:SCP}, i.e. $\hat{\bm{x}}_{k+1} + \bm{w}_k^{(i)} \notin R_j(\lambda^*_{\vert Q \vert})$, with probability one.
    \label{lemma:Q}
\end{lemma}

Intuitively, the successor state of a sample associated with an active constraint is on the boundary of the optimal solution $R_j(\lambda^*_{\vert Q \vert})$ of \cref{eq:SCP}.
Under the following \emph{non-accumulation} assumption, an active constraint exists and is unique, as long as $\vert Q \vert < N$ (i.e., not yet all samples have been discarded).
\begin{assumption}
    \label{assumption:nonAccumulation}
    Given a noise sample $\bm{w}_k \in \Delta$, the probability that the successor state $\bm{x}_{k+1}$ is on the boundary of any polytope $R_j(\lambda)$ for any $\lambda \in \R_+$ is zero.
\end{assumption}
\cref{assumption:nonAccumulation} implies that the solution to \cref{eq:SCP} is unique, and that the number of active constraints is equal to one (given $N>0$ and $\vert Q \vert < N$), as samples accumulate on the boundary of the polytope with probability zero.

\begin{figure}[t!]
	\centering
	\begin{tikzpicture}[xscale=1.8,yscale=0.97]

\fill[red!20] (-0.73,0pt) rectangle (0.73,5pt);
\fill[blue!20] (-1.15,-5pt) rectangle (1.15,0pt);

\draw[latex-latex] (-2.2,0) -- (2.2,0) ; 
\foreach \x in  {-2,-1,0,1,2} 
\draw[shift={(\x,0)},color=black] (0pt,3pt) -- (0pt,-3pt);
\foreach \x in {-2,-1,0,1,2} 
\draw[shift={(\x,0)},color=black] (0pt,0pt) -- (0pt,-3pt) node[below] 
{$\x$};

\draw (0.73,0)  node[circle,fill,inner sep=1.5pt,red] {};
\draw (-1.39,0) node[circle,fill,inner sep=1.5pt] {};
\draw (1.15,0)  node[circle,fill,inner sep=1.5pt,blue] {};
\draw (-0.41,0) node[circle,fill,inner sep=1.5pt,red] {};
\draw (0.2,0)  node[circle,fill,inner sep=1.5pt,red] {};
\draw (1.75,0)  node[circle,fill,inner sep=1.5pt] {};
\draw (-0.06,0)  node[circle,fill,inner sep=1.5pt,red] {};
\draw (-0.65,0)  node[circle,fill,inner sep=1.5pt,red] {};
\draw (-1.61,0)  node[circle,fill,inner sep=1.5pt] {};
\draw (1.33,0)  node[circle,fill,inner sep=1.5pt] {};

\draw [blue, decorate,decoration={brace,amplitude=5pt,mirror,raise=4ex}]
  (-1.15,0.1) -- (1.15,0.1) node[blue, midway,yshift=-3em]{$R_j(\lambda^*_{N_j^\text{out}-1})$};
  
\draw [red,decorate,decoration={brace,amplitude=5pt,mirror,raise=2ex}]
  (0.73,-0.05) -- (-0.73,-0.05) node[red, midway,yshift=2em]{$R_j(\lambda^*_{N_j^\text{out}})$};

\draw[thick, shift={(-0.73,0)},color=red] (0pt,5pt) -- (0pt,-0pt);
\draw[thick, shift={(0.73,0)},color=red] (0pt,5pt) -- (0pt,-0pt);

\draw[thick, shift={(-1.15,0)},color=blue] (0pt,0pt) -- (0pt,-5pt);
\draw[thick, shift={(1.15,0)},color=blue] (0pt,0pt) -- (0pt,-5pt);

\draw[shift={(-1,0)}, dashed] (0pt,0pt) -- (0pt,30pt);
\draw[shift={(1,0)}, dashed] (0pt,0pt) -- (0pt,30pt);
\draw [decorate,decoration={brace,amplitude=5pt,mirror,raise=2ex}]
  (1,17pt) -- (-1,17pt) node[midway,yshift=2em]{$R_j$};

\end{tikzpicture}
	
	\caption{
	    Bounding region $R_j = [-1, 1]$ using $N=10$ successor state samples ($N_j^\text{out} = 5$).
	    Discarding $N_j^\text{out} = 5$ samples defines the red region $R_j(\lambda^*_{N_j^\text{out}}) \subseteq R_j$, while discarding one less sample defines the blue region $R_j(\lambda^*_{N_j^\text{out}-1}) \supset R_j$.
	}
	\label{fig:Scenario1D}
\end{figure}
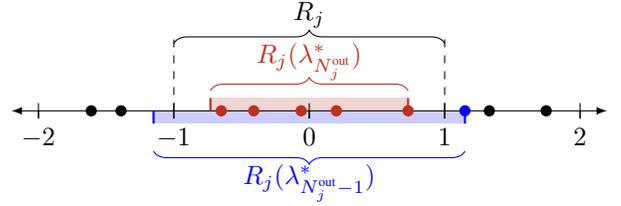

\subsubsection{Under/over-approximating regions.}

Recall from \cref{def:Nj_in} that $N_j^\text{out}$ is the number of samples leading to a successor state outside of region $R_j$.
The following lemma uses $N_j^\text{out}$ to define an under- or over-approximation of region $R_j$.

\begin{lemma}
    \label{lemma:Q_cardinality}
    When discarding $\vert Q \vert = N^\text{out}_j$ samples as per \cref{lemma:Q}, it holds that $R_j(\lambda^*_{N_j^\text{out}}) \subseteq R_j$.
    When discarding $\vert Q \vert = N^\text{out}_j - 1$ samples, it holds that $R_j(\lambda^*_{N_j^\text{out}-1}) \supset R_j$
\end{lemma}

\begin{proof}
    \cref{lemma:Q} states that at every step, we may only discard the sample of the active constraint.
    By construction, after discarding $\vert Q \vert = N_j^\text{out}$ samples, all remaining successor state samples lie within $R_j$, so $\lambda^*_{N_j^\text{out}} \leq 1$, so $R_j(\lambda^*_{N_j^\text{out}}) \subseteq R_j$.
    When we discard one less sample, i.e. $\vert Q \vert = N_j^\text{out} - 1$, we must have one sample outside of $R_j$, so $\lambda^*_{N_j^\text{out}-1} > 1$, meaning that $R_j(\lambda^*_{N_j^\text{out}-1}) \supset R_j$.
    This concludes the proof.
\end{proof}

Intuitively, $R_j(\lambda^*_{N_j^\text{out}})$ contains \emph{exactly} those samples in $R_j$, while $R_j(\lambda^*_{N_j^\text{out}-1})$ additionally contains \emph{the sample closest outside of} $R_j$, as visualized in \cref{fig:Scenario1D} for a 1D example.

\newcommand{\PxNotinR}{\ensuremath{P \big( \bm{x} \notin R_j}}
\newcommand{\PxInR}{\ensuremath{P \big( \bm{x} \in R_j}}

\subsection{Proof of \cref{theorem:Bounds}.}

The proof of \cref{theorem:Bounds} is adapted from~\citeauthor{DBLP:conf/cdc/RomaoMP20} (\citeyear[Theorem 5]{DBLP:conf/cdc/RomaoMP20}), which is based on three key assumptions: (1) the considered scenario problem belongs to the class of so-called \emph{fully-supported problems} (see~\citeauthor{DBLP:journals/siamjo/CampiG08} (\citeyear{DBLP:journals/siamjo/CampiG08}) for a definition), (2) its solution is unique, and (3) discarded samples violate the optimal solution with probability one.
In our case, requirement (2) is implied by \cref{assumption:nonAccumulation}, (3) is satisfied by \cref{lemma:Q}, and the problem is fully-supported because the number of decision variables is one.
Under these requirements,~\citeauthor{DBLP:conf/cdc/RomaoMP20} (\citeyear[Theorem 5]{DBLP:conf/cdc/RomaoMP20}) states that the risk associated with an optimal solution $\lambda^*_{\vert Q \vert}$ to \cref{eq:SCP} for $\vert Q \vert$ discarded samples satisfies the following expression:
\begin{equation}
\begin{split}
    \amsmathbb{P}^N \Big\{ & \PxNotinR (\lambda^*_{\vert Q \vert}) \big) \leq \epsilon \Big\}
    \\
    & \qquad = 1 - \sum_{i=0}^{\vert Q \vert} \binom Ni \epsilon^i (1-\epsilon)^{N-i},
\end{split}
\label{eq:Proof1}
\end{equation}
where we omit the subscripts in $P_{\bm{w}_k}(\cdot)$ and $\bm{x}_{k+1}$ for brevity.
\cref{eq:Proof1} is the cumulative distribution function (CDF) of a beta distribution with parameters $\vert Q \vert + 1$ and $N-\vert Q \vert$.
We denote this CDF by $F_{\vert Q \vert}(\epsilon)$, where we explicitly write the dependency on $\vert Q \vert$.
Hence, we obtain
\begin{equation}
    F_{\vert Q \vert}(\epsilon) = \amsmathbb{P}^N \Big\{ \PxNotinR (\lambda^*_{\vert Q \vert}) \big) \leq \epsilon \Big\} = \tilde{\beta}.
\label{eq:Proof1b}
\end{equation}
Thus, if we discard $\vert Q \vert$ samples, \cref{eq:Proof1b} returns the confidence probability $\tilde{\beta} \in (0,1)$ by which the probability of $\bm{x}_{k+1} \notin R_j(\lambda^*_{\vert Q \vert})$ is upper bounded by $\epsilon$.
Conversely, we can compute the value of $\epsilon$ needed to obtain a confidence probability of $\tilde{\beta}$, using the percent point function (PPF) $G_{\vert Q \vert}(\tilde{\beta})$:
\begin{equation}
    \amsmathbb{P}^N \Big\{ \PxNotinR (\lambda^*_{\vert Q \vert}) \big) \leq G_{\vert Q \vert}(\tilde{\beta}) \Big\} = \tilde{\beta}.
    \label{eq:Proof2}
\end{equation}
The PPF is the inverse of the CDF, so by definition, we have
\begin{equation}
    \epsilon = G_{\vert Q \vert}(\tilde{\beta}) = G_{\vert Q \vert}\big( F_{\vert Q \vert} (\epsilon) \big).
    \label{eq:Proof2b}
\end{equation}
Note that $P(\bm{x} \in R) + P(\bm{x} \notin R) = 1$, so \cref{eq:Proof1b} equals
\begin{equation}
    F_{\vert Q \vert}(\epsilon) = \amsmathbb{P}^N \Big\{ \PxInR (\lambda^*_{\vert Q \vert}) \big) \geq 1-\epsilon \Big\} = \tilde{\beta}.
\label{eq:Proof3}
\end{equation}
By defining $p = 1 - \epsilon$, \cref{eq:Proof3,eq:Proof2b} are combined as
\begin{equation}
\begin{split}
    \amsmathbb{P}^N \Big\{ 1 - G_{\vert Q \vert}(\tilde{\beta}) \leq \PxInR (\lambda^*_{\vert Q \vert}) \big) \Big\}&
    \\
    = 1 - \sum_{i=0}^{\vert Q \vert} \binom Ni (1-p)^i p^{N-i} = \tilde{\beta}.&
    \label{eq:Proof4}
\end{split}
\end{equation}
In what follows, we separately use \cref{eq:Proof3} to prove the lower and upper bound of the probability interval in \cref{theorem:Bounds}.

\paragraph{Lower bound.}
There are $N$ possible values for $\vert Q \vert$, ranging from $0$ to $N-1$.
The case $\vert Q \vert = N$ (i.e. all samples are discarded) is treated as a special case in \cref{theorem:Bounds}.
We fix $\tilde{\beta} = 1 - \frac{\beta}{2 N}$ in \cref{eq:Proof4}, yielding the series of equations
{\small
\begin{alignat}{3}
    & \amsmathbb{P}^N \Big\{ 1 - G_{0} \big( 1 - \frac{\beta}{2 N} \big) \leq \PxInR (\lambda^*_{0}) \big) \Big\} && = && 1 - \frac{\beta}{2 N}
    \nonumber
    \\
    & \qquad\qquad\qquad\qquad \vdots && && \vdots
    \label{eq:Proof_sysEquations_low}
    \\
    & \amsmathbb{P}^N \Big\{ 1 - G_{N-1} \big( 1 - \frac{\beta}{2 N} \big) \leq \PxInR (\lambda^*_{N-1}) \big) \Big\} && = && 1 - \frac{\beta}{2 N}.
    \nonumber
\end{alignat}
}
Denote the event that $1 - G_{n} \big( 1 - \frac{\beta}{2 N} \big) \leq \PxInR (\lambda^*_{n}) \big)$ for $n = 0,\ldots,N-1$ by $\mathcal{A}_n$.
Regardless of $n$, this event has a probability of $\amsmathbb{P}^N \{ \mathcal{A}_n \} = 1 - \frac{\beta}{2N}$, and its complement $\mathcal{A}_n'$ of $\amsmathbb{P}^N \{ \mathcal{A}_n' \} = \frac{\beta}{2N}$.
Via Boole's inequality, we know that 
\begin{equation}
    \amsmathbb{P}^N \Big\{ \bigcup_{i = 0}^{N-1} \mathcal{A}_n' \Big\} \leq 
    \sum_{i=0}^{N-1} \amsmathbb{P}^N \big\{ \mathcal{A}_n' \big\}
    =
    \frac{\beta}{2N} N
    =
    \frac{\beta}{2}.
    \label{eq:UnionBound}
\end{equation}
Thus, for the intersection of all events in \cref{eq:Proof_sysEquations_low} we have
\begin{equation}
    \amsmathbb{P}^N \Big\{ \bigcap_{i = 0}^{N-1} \mathcal{A}_n \Big\} 
    = 
    1 - \amsmathbb{P}^N \Big\{ \bigcup_{i = 0}^{N-1} \mathcal{A}_n' \Big\}
    \geq 1 - \frac{\beta}{2}.
    \label{eq:UnionBound2}
\end{equation}
After observing the samples at hand, we replace $\vert Q \vert$ by the actual value of $N_j^\text{out}$ (as per \cref{def:Nj_in}), giving one of the expression in \cref{eq:Proof_sysEquations_low}.
The probability that this expression holds cannot be smaller than that of the intersection of all events in \cref{eq:UnionBound2}. 
Thus, we obtain
\begin{equation}
    \amsmathbb{P}^N \Big\{
        \munderbar{p} 
        \leq 
        \PxInR (\lambda^*_{N_j^\text{out}}) \big)
    \Big\} \geq 1 - \frac{\beta}{2},
    \label{eq:Proof6_lowerBound}
\end{equation}
where $\munderbar{p} = 0$ if $N_j^\text{out} = N$ (which trivially holds with probability one), and otherwise $\munderbar{p} = 1 - G_{N_j^\text{out}} (1 - \frac{\beta}{2 N})$ is the solution for $p$ to \cref{eq:Proof4}, with $\vert Q \vert = N_j^\text{out}$ and $\tilde{\beta} = 1-\frac{\beta}{2 N}$:
\begin{equation}
\begin{split}
    1 - \frac{\beta}{2 N} &= 1 - \sum_{i=0}^{N_j^\text{out}} \binom Ni (1-p)^i p^{N-i},
    \label{eq:Proof7}
\end{split}
\end{equation}
which is equivalent to \cref{eq:TheoremLowerBound}.

\subsubsection{Upper bound.}
\cref{eq:Proof4} is rewritten as an upper bound as
\begin{equation}
    \amsmathbb{P}^N \Big\{ \PxInR (\lambda^*_{\vert Q \vert}) \big) < 1 - G_{\vert Q \vert}(\tilde{\beta}) \Big\} = 1 - \tilde{\beta},
    \label{eq:Proof8}
\end{equation}
where again, $\vert Q \vert$ can range from $0$ to $N-1$.
However, to obtain high-confidence guarantees on the upper bound, we now fix $\tilde{\beta} = \frac{\beta}{2N}$, which yields the series of equations
{\small
\begin{alignat}{3}
    & \amsmathbb{P}^N \Big\{ \PxInR (\lambda^*_{0}) \big) < 1 - G_{0} \big( \frac{\beta}{2 N} \big) \Big\} & = & 1 - \frac{\beta}{2 N}
    \nonumber
    \\
    & \qquad\qquad\qquad\qquad \vdots & & \vdots
    \label{eq:Proof_sysEquations_upp}
    \\
    & \amsmathbb{P}^N \Big\{ \PxInR (\lambda^*_{N-1}) \big) < 1 - G_{N-1} \big( \frac{\beta}{2 N} \big) \Big\} & = & 1 - \frac{\beta}{2 N}.
    \nonumber
\end{alignat}
}
Analogous to the lower bound case, Boole's inequality implies that the intersection of all expressions in \cref{eq:Proof_sysEquations_upp} has a probability of at least $1 - \frac{\beta}{2}$. 
After observing the samples at hand, we replace $\vert Q \vert$ by $N_j^\text{out} - 1$, yielding one of the expressions in \cref{eq:Proof_sysEquations_upp}.
For this expression, it holds that
\begin{equation}
    \amsmathbb{P}^N \Big\{
        \PxInR (\lambda^*_{N_j^\text{out} - 1}) \big) \leq \bar{p}
    \Big\} \geq 1 - \frac{\beta}{2},
    \label{eq:Proof10_upperBound}
\end{equation}
where $\bar{p} = 1$ if $N_j^\text{out} = 0$ (which trivially holds with probability one), and otherwise $\bar{p} = 1 - G_{N_j^\text{out} - 1}(\frac{\beta}{2 N})$ is the solution for $p$ to \cref{eq:Proof4}, with $\vert Q \vert = N_j^\text{out} - 1$ and $\tilde{\beta} = \frac{\beta}{2 N}$:
\begin{equation}
    \frac{\beta}{2 N} = 1 - \sum_{i=0}^{N_j^\text{out} -1 } \binom Ni (1-p)^i p^{N-i},
    \label{eq:Proof11}
\end{equation}
which is equivalent to \cref{eq:TheoremUpperBound}.

\subsubsection{Probability interval.}
We invoke \cref{lemma:Q_cardinality}, which states that $R_j(\lambda^*_{N_j^\text{out}}) \subseteq R_j \subset R_j(\lambda^*_{N_j^\text{out} - 1})$, so we have
\begin{equation}
\begin{split}
    \PxInR (\lambda^*_{N_j^\text{out}}) \big)
    & \leq 
    P (\bm{x} \in R_j)
    =
    P(s_i, a_l)(s_j)
    \label{eq:Proof12}
    \\
    & < \PxInR (\lambda^*_{N_j^\text{out} - 1}) \big).
\end{split}
\end{equation}
We use \cref{eq:Proof12} to write \cref{eq:Proof6_lowerBound,eq:Proof10_upperBound} in terms of the transition probability $P(s_i, a_l)(s_j)$.
Finally, by applying Boole's inequality, we combine \cref{eq:Proof6_lowerBound,eq:Proof10_upperBound} as follows:
\begin{equation}
\begin{split}
    & \amsmathbb{P}^N \Big\{
        \munderbar{p} \leq P(s_i, a_l)(s_j) \, \bigcap \, P(s_i, a_l)(s_j) \leq \bar{p}
    \Big\} 
    \\
    & = \amsmathbb{P}^N \Big\{
        \munderbar{p} \leq P(s_i, a_l)(s_j) \leq \bar{p}
    \Big\} 
    \geq 1 - \beta,
    \label{eq:Proof13_interval}
\end{split}
\end{equation}
which is equal to \cref{eq:TheoremBounds}, so we conclude the proof.

\subsection{Example use of \cref{theorem:Bounds}}

\label{example:ScenarioApproach}

We provide an example to illustrate how the number of samples $N$ affects the tightness of the probability intervals returned by \cref{theorem:Bounds}.
Consider a system with a 1-dimensional state $\bm{x}_k \in \R$, where the probability density function $p_{\bm{w}_k}(\bm{x}_{k+1} \ | \ \hat{\bm{x}}_{k+1} = \bm{d}_l)$ for a specific action $a_l$ with target point $d_l$ is given by a uniform distribution over the domain $[-4, 4]$.
For a given region $R_j = [-1,1]$ (which is also shown in \cref{fig:Scenario1D}), we want to evaluate the probability that $\bm{x}_{k+1} \in R_j$, which is clearly $0.25$. 
To this end, we apply \cref{theorem:Bounds} for different numbers of samples $N \in [25, 12800]$, and a confidence level of $\beta=0.1$ or $\beta=0.01$.
Since the obtained probability bounds are random variables through their dependence on the set of $N$ samples of the noise, we repeat the experiment $100,000$ times for every value of $N$.
The resulting bounds on the transition probability are shown in \cref{fig:empiricalValidation}.
The plot shows that the uncertainty in the transition probability is reduced by increasing the number of samples.

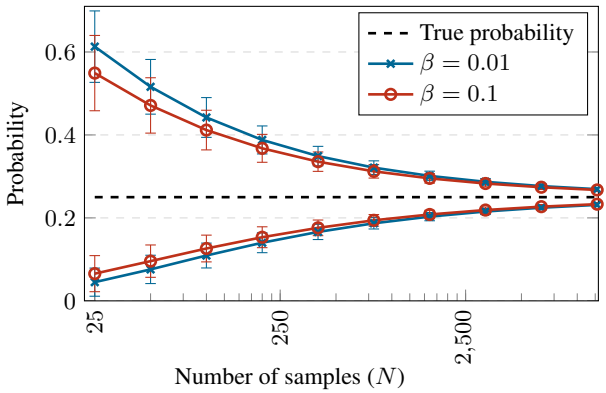
\begin{figure}[t!]
	\centering
	    \footnotesize\begin{tikzpicture}
      \begin{axis}[
          width=\linewidth,
          height=5.5cm,
          ymajorgrids,
          grid style={dashed,gray!30},
          xlabel=Number of samples ($N$),
          ylabel=Probability,
          x label style={at={(axis description cs:0.4,-0.2)}},
          x tick label style={rotate=90,anchor=east},
          xmode=log,
          log ticks with fixed point,
          xmin=22,
          xmax=13000,
          ymin=0,
          ymax=0.71,
          xtick={25,250,2500,25000},
          xtick pos=left,
          ytick={0,0.2,0.4,0.6},
          every axis plot/.append style={line width=1pt},
          legend cell align={left},
          legend columns=1,
        ]

        \addplot+[
          black,
          dashed,
          mark=none
        ]
        coordinates {
        (25, 0.25)
        (12800, 0.25)
        };
        
        \addplot+[
          mark=x,
          color=MidnightBlue,
          error bars/.cd,
            y dir=both,
            y explicit,
        ]
        coordinates {
        (25, 0.0451031) +- (0,0.0339226)
        (50, 0.0758369) +- (0,0.0341477)
        (100, 0.109287) +- (0,0.0299612)
        (200, 0.140166) +- (0,0.0240263)
        (400, 0.166169) +- (0,0.0183472)
        (800, 0.186985) +- (0,0.0136881)
        (1600, 0.203126) +- (0,0.0100159)
        (3200, 0.215401) +- (0,0.0072561)
        (6400, 0.224562) +- (0,0.00521526)
        (12800, 0.23138561) += (0,0.00371907)
        };
        
        \addplot+[
          mark=o,
          color=red,
          error bars/.cd,
            y dir=both,
            y explicit,
        ]
        coordinates {
        (25, 0.065582) +- (0,0.0434891)
        (50, 0.095761) +- (0,0.038991)
        (100, 0.126207) +- (0,0.0322542)
        (200, 0.153394) +- (0,0.0251742)
        (400, 0.176011) +- (0,0.0188726)
        (800, 0.194169) +- (0,0.0139199)
        (1600, 0.208229) +- (0,0.0101429)
        (3200, 0.218945) +- (0,0.0072842)
        (6400, 0.22706) +- (0,0.00524142)
        (12800, 0.23310191) +- (0,0.00374732)
        };
        
        \addplot+[
          mark=x,
          color=MidnightBlue,
          error bars/.cd,
            y dir=both,
            y explicit,
        ]
        coordinates {
        (25, 0.612857) +- (0,0.086119)
        (50, 0.516028) +- (0,0.0659096)
        (100, 0.442108) +- (0,0.0479324)
        (200, 0.387955) +- (0,0.0337365)
        (400, 0.349004) +- (0,0.0234804)
        (800, 0.321158) +- (0,0.0163829)
        (1600, 0.301241) +- (0,0.0114166)
        (3200, 0.286971) +- (0,0.00798212)
        (6400, 0.276666) +- (0,0.00559007)
        (12800, 0.26925964) +- (0,0.0039115)
        };
        
        \addplot+[
          mark=o,
          color=red,
          error bars/.cd,
            y dir=both,
            y explicit,
        ]
        coordinates {
        (25, 0.549077) +- (0,0.0907803)
        (50, 0.471126) +- (0,0.0667335)
        (100, 0.411734) +- (0,0.0477994)
        (200, 0.367734) +- (0,0.0336444)
        (400, 0.335569) +- (0,0.0233622)
        (800, 0.312283) +- (0,0.0162941)
        (1600, 0.295294) +- (0,0.0113888)
        (3200, 0.282917) +- (0,0.00793146)
        (6400, 0.273943) +- (0,0.00557889)
        (12800, 0.26738347) +- (0,0.00392236)
        };

        \legend{{True probability},{$\beta = 0.01$},{$\beta = 0.1$}}
      \end{axis}
    \end{tikzpicture}
	\caption{Probability intervals (with standard deviations),
	obtained from \cref{theorem:Bounds} with $\beta=0.01$ or $\beta=0.1$ on a transition with a true probability of $0.25$ (note the log scale).}
	\label{fig:empiricalValidation}
\end{figure}
    \section{Details on UAV Benchmark}
\label{appendix:UAV}

\subsection{Explicit model formulation}
\label{subappendix:UAV_model}

The 6-dimensional state vector of the \gls{UAV} model is $\bm{x} = [p_x, v_x, p_y, v_y, p_z, v_z]^\top \in \R^6$, where $p_i$ and $v_i$ are the position and velocity in the direction $i$.
Every element of the vector of control inputs $\bm{u} = [f_x, f_y, f_z]^\top \in \R^3$ is constrained to the interval $[-4, 4]$.
The discrete-time dynamics are an extension of the lower-dimensional case in~\citeauthor{Badings2021FilterUncertainty} (\citeyear{Badings2021FilterUncertainty}), and are written in the form of \cref{eq:continuousSystem} as follows:
\begin{equation}
    \bm{x}_{k+1} = \scalemath{0.7}{	\begin{bmatrix}
        1 & 1 & 0 & 0 & 0 & 0 \\
        0 & 1 & 0 & 0 & 0 & 0 \\
        0 & 0 & 1 & 1 & 0 & 0 \\
        0 & 0 & 0 & 1 & 0 & 0 \\
        0 & 0 & 0 & 0 & 1 & 1 \\
        0 & 0 & 0 & 0 & 0 & 1
    \end{bmatrix}}
    \bm{x}_{k} + \scalemath{0.7}{	\begin{bmatrix}
        0.5 & 0 & 0 \\
        1 & 0 & 0 \\
        0 & 0.5 & 0 \\
        0 & 1 & 0 \\
        0 & 0 & 0.5 \\
        0 & 0 & 1
    \end{bmatrix}} \bm{u}_{k} + \bm{w}_k,
    \label{eq:UAV_explicitModel}
\end{equation}
where $\bm{w}_k$ is the effect of turbulence on the continuous state, which we model using a Dryden gust model; we use a Python implementation by~\citeauthor{bohn2019deep} (\citeyear{bohn2019deep}).
Note that the drift term $\bm{q}_k = 0$, and is thus omitted from  \cref{eq:UAV_explicitModel}.
Since the model in \cref{eq:UAV_explicitModel} is not fully actuated (it has only $3$ controls, versus a state of dimension $6$), we group every two discrete time steps together and rewrite the model as follows:
\begin{equation}
    \bm{x}_{k+2} = \bar{A} \bm{x}_k + \bar{U} \bm{u}_{k,k+1} + \bm{w}_{k,k+1},
    \label{eq:UAV_fullyActuated}
\end{equation}
where $\bar{A} \in \R^{6 \times 6}$ and $\bar{U} \in \R^{6 \times 6}$ are properly redefined matrices, and $\bm{u}_{k,k+1} \in \R^6$ and $\bm{w}_{k,k+1} \in \R^6$ reflect the control inputs and process noise at both time steps $k$ and $k+1$, combined in one vector.

The objective of this problem is to reach the goal region, which is also depicted in \cref{fig:UAV_layout}, within $64$ steps (i.e. $32$ steps with the model in \cref{eq:UAV_fullyActuated}).
This goal region is written as the following set of continuous states (for brevity, we omit an explicit definition of the critical regions):
\begin{equation}
\begin{split}
    \goalRegion = \{ \bm{x} \in \R^6 \ \vert \ &11 \leq p_x \leq 15, \\
    \ \ &1 \leq p_y \leq 5, \ \ -7 \leq p_z \leq -3 \}.
\end{split}
\end{equation}

\subsection{Run times and size of model abstraction}

\label{subappendix:UAV_results}

The average run times and \gls{iMDP} model sizes (over 10 runs) for all iterations of the \gls{UAV} benchmark are presented in \cref{tab:DetailedResults}.
Computing the states and actions took $\SI{15}{\minute}$, but since this step is only performed in the first iteration, we omit the value from the table.
The number of states (which equals the number of regions in the partition) and choices (the total number of state-action pairs) of the \gls{iMDP} are independent of the value of $N$.
By contrast, the number of transitions increases with $N$, because the additional noise samples may reveal more possible outcomes of a state-action pair.

\begin{table*}[t!]

\caption{Run times per iteration (which excludes the computation of states and actions) and model sizes for the \gls{UAV} (under high turbulence), and 2-zone and 1-zone building temperature regulation (BAS) benchmarks.}
\label{tab:DetailedResults}

\small{
\begin{tabularx}{\textwidth}{@{} lX|lXl|XXX @{} } \toprule
    \multicolumn{2}{c|}{\textbf{Instance}} & \multicolumn{3}{c|}{\textbf{Run times (per iteration, in seconds)}} & \multicolumn{3}{c}{\textbf{\gls{iMDP} model size}}
    \\
    Benchmark & Samples $N$ & {Update intervals} & Write \gls{iMDP} & Model checking & States & Choices & Transitions
    \\ \midrule
    UAV & 25   & 11.4  & 29.0  & 39.2  & 25,516 & 1,228,749 & 9,477,795 \\
    UAV & 50   & 15.4  & 34.4  & 46.0  & 25,516 & 1,228,749 & 11,350,692 \\
    UAV & 100  & 18.6  & 39.3  & 51.8  & 25,516 & 1,228,749 & 13,108,282 \\
    UAV & 200  & 24.3  & 44.4  & 59.1  & 25,516 & 1,228,749 & 14,921,925 \\
    UAV & 400  & 35.4  & 50.6  & 68.5  & 25,516 & 1,228,749 & 17,203,829 \\
    UAV & 800  & 55.4  & 58.3  & 80.2  & 25,516 & 1,228,749 & 19,968,663 \\
    UAV & 1600 & 92.2  & 64.9  & 90.6  & 25,516 & 1,228,749 & 22,419,161 \\
    UAV & 3200 & 164.1 & 68.6  & 94.4  & 25,516 & 1,228,749 & 23,691,107 \\
    UAV & 6400 & 312.1 & 69.8  & 95.5  & 25,516 & 1,228,749 & 23,958,183 \\
    UAV & 12800& 611.2 & 69.8  & 95.1  & 25,516 & 1,228,749 & 23,969,028 \\ \midrule
    BAS (2-zone) & 25   & 13.9  & 73.6  & 105.9  & 35,722 & 1,611,162 & 24,929,617 \\
    BAS (2-zone) & 50   & 43.2  & 107.7 & 160.5  & 35,722 & 1,611,162 & 37,692,421 \\
    BAS (2-zone) & 100  & 50.8  & 153.2 & 230.2  & 35,722 & 1,611,162 & 53,815,543 \\
    BAS (2-zone) & 200  & 59.1  & 205.8 & 315.2  & 35,722 & 1,611,162 & 72,588,298 \\
    BAS (2-zone) & 400  & 72.3  & 260.6 & 423.2  & 35,722 & 1,611,162 & 91,960,970 \\
    BAS (2-zone) & 800  & 96.2  & 316.8 & 504.3  & 35,722 & 1,611,162 & 112,028,449 \\
    BAS (2-zone) & 1600 & 132.6 & 370.0 & 614.8  & 35,722 & 1,611,162 & 131,844,146 \\
    BAS (2-zone) & 3200 & 200.1 & 432.9 & 789.4  & 35,722 & 1,611,162 & 154,155,445 \\
    BAS (2-zone) & 6400 & 331.8 & 513.2 & 982.2  & 35,722 & 1,611,162 & 182,175,229 \\
    BAS (2-zone) & 12800& 545.5 & 601.8 & 1177.8 & 35,722 & 1,611,162 & 218,031,384 \\ \midrule
    BAS (1-zone) & 25   & 1.62 & 0.06 & 1.39 & 381 & 1,511 & 20,494 \\
    BAS (1-zone) & 50   & 1.81 & 0.08 & 1.39 & 381 & 1,511 & 27,327 \\
    BAS (1-zone) & 100  & 1.87 & 0.09 & 1.40 & 381 & 1,511 & 33,871 \\
    BAS (1-zone) & 200  & 1.96 & 0.11 & 1.40 & 381 & 1,511 & 40,368 \\
    BAS (1-zone) & 400  & 2.06 & 0.13 & 1.42 & 381 & 1,511 & 47,333 \\
    BAS (1-zone) & 800  & 2.26 & 0.14 & 1.40 & 381 & 1,511 & 54,155 \\
    BAS (1-zone) & 1600 & 2.59 & 0.16 & 1.39 & 381 & 1,511 & 59,686 \\
    BAS (1-zone) & 3200 & 3.24 & 0.17 & 1.40 & 381 & 1,511 & 65,122 \\
    BAS (1-zone) & 6400 & 4.50 & 0.18 & 1.40 & 381 & 1,511 & 70,245 \\
    BAS (1-zone) & 12800& 7.01 & 0.20 & 1.38 & 381 & 1,511 & 76,076 \\ \bottomrule
    
\end{tabularx}
}

\end{table*}

\section{Details on Building Temperature Regulation Benchmark}

\label{appendix:BAS}

\subsection{Explicit model formulation}

\label{subappendix:BAS_model}

This model is a variant of the two-zone building automation system benchmark with stochastic dynamics in~\citeauthor{DBLP:journals/corr/abs-1803-06315} (\citeyear{DBLP:journals/corr/abs-1803-06315}).
The state vector of the model is $\bm{x} = [T^z_1, T^z_2, T^r_1, T^r_2]^\top \in \R^4$, reflecting both room (zone) temperatures ($T^z$) and both radiator temperatures ($T^r$).
The control inputs $\bm{u} = [T^\text{ac}_1, T^\text{ac}_2, T^\text{boil}_1, T^\text{boil}_2]^\top \in \R^4$ are the air conditioning (ac) and boiler temperatures, respectively, which are constrained within $14 \leq T^\text{ac} \leq 26$ and $65 \leq T^\text{boil} \leq 85$.
The changes in the temperature of both zones are governed by the following thermodynamics:
\begin{subequations}
\begin{align}
    \dot{T}^z_1 =& \frac{1}{C_1} \Big[ \frac{T^z_2 - T^z_1}{R_{1,2}} + \frac{T_\text{wall} - T^z_1}{R_{1,\text{wall}}} 
    \\
    &+ m C_{pa} (T^\text{ac}_1 - T^z_1) + P_\text{out} (T_1^r - T_1^z) \Big]
    \nonumber
    \\
    \dot{T}^z_2 =& \frac{1}{C_2} \Big[ \frac{T^z_1 - T^z_2}{R_{1,2}} + \frac{T_\text{wall} - T^z_2}{R_{2,\text{wall}}} 
    \\
    &+ m C_{pa} (T^\text{ac}_2 - T^z_2) + P_\text{out} (T_2^r - T_2^z) \Big],
    \nonumber
\end{align}
\label{eq:BAS_zoneTemps}
\end{subequations}
where $C_i$ is the thermal capacitance of zone $i$, and $R_{i,j}$ is the resistance between zones $i$ and $j$, $m$ is the air mass flow, $C_{pa}$ is the specific heat capacity of air, and $P_\text{out}$ is the rated output of the radiator.
Similarly, the dynamics of the radiator in room $i$ are governed by the following equation:
\begin{equation}
    \dot{T}_i^r = k_1 (T_i^z - T_i^r) + k_0 w (T_i^\text{boil} - T_i^r), 
    \label{eq:BAS_radiatorTemp}
\end{equation}
where $k_0$ and $k_1$ are constant parameters, and $w$ is the water mass flow from the boiler.
For the precise parameter values used, we refer to our codes, which are provided in the supplementary material.
By discretizing \cref{eq:BAS_zoneTemps,eq:BAS_radiatorTemp} by a forward Euler method at a time resolution of $\SI{15}{\minute}$, we obtain the following model in explicit form:
\begin{align}
    \bm{x}_{k+1} = & \scalemath{0.8}{\begin{bmatrix}
        0.8425 & 0.0537 & -0.0084 &  0.0000 \\
        0.0515 & 0.8435 &  0.0000 & -0.0064 \\
        0.0668 & 0.0000 &  0.8971 &  0.0000 \\
        0.0000 & 0.0668 &  0.0000 &  0.8971
    \end{bmatrix}} \bm{x}_k
    \nonumber
    \\ &
    + \scalemath{0.8}{\begin{bmatrix}
        0.0584 & 0.0000 & 0.0000 & 0.0000 \\
        0.0000 & 0.0599 & 0.0000 & 0.0000 \\
        0.0000 & 0.0000 & 0.0362 & 0.0000 \\
        0.0000 & 0.0000 & 0.0000 & 0.0362
    \end{bmatrix}} \bm{u}_k 
    \\ & 
    + \scalemath{0.8}{\begin{bmatrix}
        1.2291 \\ 1.0749 \\ 0.0000 \\ 0.0000
    \end{bmatrix}} + \Gauss[\scalemath{0.8}{\begin{bmatrix}0 \\ 0 \\ 0 \\ 0 \end{bmatrix}}]{\scalemath{0.8}{\begin{bmatrix}
        0.01 & 0 & 0 & 0 \\
        0 & 0.01 & 0 & 0 \\
        0 & 0 & 0.01 & 0 \\
        0 & 0 & 0 & 0.01
    \end{bmatrix}}},
    \nonumber
\end{align}
where $\Gauss[\bm\mean]{\cov} \in \R^n$ is a multivariate Gaussian random variable with mean $\bm\mean \in \R^n$ and covariance matrix $\cov$. 

The goal in this problem is to reach a temperature of $19.9 - \SI{20.1}{\celsius}$ in both zones within $32$ discrete time steps of $\SI{15}{\minute}$.
As such, the goal region $\goalRegion$ and critical region $\criticalRegion$ are:
\begin{alignat}{2}
    \goalRegion &= \{ \bm{x} \in \R^2 \ \vert \ &&19.9 \leq T^z_1 \leq 20.1, \ 
    \nonumber
    \\
    & &&19.9 \leq T^z_2 \leq 20.1 \}
    \\
    \criticalRegion &= \varnothing. &&
    \nonumber
\end{alignat}

\subsection{Run times and size of model abstraction}

\label{subappendix:BAS_results}

The run times and \gls{iMDP} model sizes for all iterations of the 2-zone building temperature regulation benchmark are presented in \cref{tab:DetailedResults}.
Computing the states and actions took 5:25 $\SI{}{\minute}$, but we omit this value from the table as this step is only performed in the first iteration.
The discussion of model sizes is analogous to \cref{subappendix:UAV_results} on the \gls{UAV} benchmark.
    \section{Benchmarks against other tools}

\label{appendix:Benchmarks}

\subsection{Benchmark against \stochy}
\label{appendix:StocHy}
We provide a comparison between our approach and \stochy~\citep{DBLP:conf/tacas/CauchiA19} on a reduced version of the temperature regulation problem in \cref{subappendix:BAS_model} with only one zone.
Thus, the state vector becomes $\bm{x} = [T^z, T^r]^\top$, and the control inputs are $\bm{u} = [T^{\text{ac}}, T^{\text{boil}}]^\top$, which are constrained to $14 \leq T^\text{ac} \leq 28$ and $-10 \leq T^\text{boil} \leq 10$ (note that $T^\text{boil}$ is now relative to the nominal boiler temperature of $\SI{75}{\celsius}$).
Using the same dynamics as in \cref{eq:BAS_zoneTemps,eq:BAS_radiatorTemp}, we obtain the following model in explicit form:
\begin{align}
    \bm{x}_{k+1} =& \begin{bmatrix}
        0.8820 & 0.0058 \\
        0.0134 & 0.9625 
    \end{bmatrix} \bm{x}_k + \begin{bmatrix}
        0.0584 & 0.0000 \\
        0.0000 & 0.0241
    \end{bmatrix} \bm{u}_k 
    \nonumber
    \\
    &+ \begin{bmatrix}
        0.9604 \\
        1.3269
    \end{bmatrix} + \Gauss[\begin{bmatrix} 0 \\ 0 \end{bmatrix}]{\begin{bmatrix}
        0.02 & 0 \\
        0 & 0.1
    \end{bmatrix}}. 
\end{align}
The objective in this problem is to reach a zone temperature of $20.9 - \SI{21.1}{\celsius}$ within $64$ discrete time steps of $\SI{15}{\minute}$.
As such, the goal region $\goalRegion$ and critical region $\criticalRegion$ are:
\begin{equation}
    \goalRegion = \{ \bm{x} \in \R^2 \ \vert \ 20.9 \leq T^z \leq 21.1 \}, \quad \criticalRegion = \varnothing.
\end{equation}
We partition the continuous state space in a rectangular grid of $19 \times 20$ regions of width $\SI{0.2}{\celsius}$, which is centered at $T^z = \SI{21}{\celsius}$ and $T^r = \SI{38}{\celsius}$.
As such, the partition covers zone temperatures between $19.1 - \SI{22.9}{\celsius}$, and radiator temperatures between $36 - \SI{40}{\celsius}$.

\subsubsection{Abstraction method.}
Similar to our approach, \stochy generates robust \gls{iMDP} abstractions of the model above.
However, while our approach also works when the distribution of the noise is unknown, \stochy requires precise knowledge of the noise distribution.
Another difference is that \stochy discretizes the control inputs of the dynamical system, to obtain a finite action space.
An \gls{iMDP} action is then associated with the execution of a specific value of the control input $\bm{u}_k$.
In our approach, an \gls{iMDP} action instead reflects an attempted transition to a given target point, and the actual control input $\bm{u}_k$ (calculated using the control law \cref{eq:controlLaw}) depends on the precise continuous state $\bm{x}_k$.

\begin{table*}[t!]

\caption{Reachability probability, the average simulated reachability probability, and run times for our method and \sreachtools, under the default (x1) and increased (x10) noise covariance.}
\label{tab:SReachTools}

\small{
\begin{tabularx}{\textwidth}{@{} l|lXXXl|XX @{} } \toprule
    \multicolumn{1}{c|}{}
    & 
    \multicolumn{5}{c|}{\textbf{Our method}} 
    & 
    \multicolumn{2}{c}{\textbf{\sreachtools}}
    \\
    & Initial abstraction & $N=25$ & $N=100$ & $N=400$ & $N=1600$ & Particle & Voronoi \\ \midrule
    \textbf{Noise covariance x1} & & & & & & & \\
    Reachability probability    & ---   & 0.44  & 0.81  & 0.95   & 0.98   & 0.88  & 0.85 \\ 
    Simulated probability       & ---   & 1.00  & 1.00  & 1.00   & 1.00   & 0.83  & 0.86 \\ 
    Run time (s)    & 3.420 & 6.966 & 9.214 & 11.016 & 17.294 & 0.703 & 3.462 \\ \midrule
    \textbf{Noise covariance x10} & & & & & & & \\
    Reachability probability    & ---   & 0.19   & 0.36   & 0.52   & 0.62   & 0.40  & NaN \\ 
    Simulated probability       & ---   & 0.59   & 0.63   & 0.59   & 0.65   & 0.19  & NaN \\ 
    Run time (s)    & 3.372 & 11.433 & 15.909 & 18.671 & 25.452 & 2.495 & NaN \\ \bottomrule
    
\end{tabularx}
}

\end{table*}

\subsubsection{Our approach is faster.}
The run times and model sizes of our approach are reported in \cref{tab:DetailedResults} under BAS (1-zone).
For our approach, the time to compute the states and actions (needed only in the first iteration) is $\SI{0.1}{\second}$, and run times per iteration (excluding verification times) vary from $1.7-\SI{7.2}{\second}$, depending on the value of $N$.
Notably, the \stochy run times are \emph{orders of magnitude higher}: the abstraction procedure of \stochy (excluding verification time) took 45:28 $\SI{}{\minute}$, compared to $\SI{7.2}{\second}$ for our approach with the maximum of $N=12800$ samples (as reported in \cref{tab:DetailedResults}).
We omit a comparison of the verification times, since \stochy does not leverage \prism like our method does.

These results provide a clear message: our approach is more general and significantly faster, and (as shown earlier in \cref{fig:BAS_stochy}) generates results that are qualitatively similar to those obtained from \stochy.

\subsection{Benchmark against \sreachtools}

\label{appendix:SReachTools}

We benchmark our approach against the spacecraft rendezvous problem supplied with the \matlab toolbox \sreachtools~\citep{DBLP:conf/hybrid/VinodGO19}.
In this problem, one spacecraft must dock with another within $8$ time steps, while remaining in a target tube. 
The goal is to compute a controller that maximizes the probability to achieve this objective. 
The 4-dimensional state $\bm{x} = [p_x, p_x, v_x, v_y]^\top \in \R^4$ describes the position and velocity in both directions.
We adopt the same discrete-time dynamics as used in~\citeauthor{DBLP:conf/hybrid/VinodGO19} (\citeyear{DBLP:conf/hybrid/VinodGO19}) and assume the same Gaussian noise (see the reference for details).

This problem is a finite-horizon reach-avoid problem with a rectangular goal region (target set), while the safe set is a convex tube, as also shown in  \cref{fig:SReachTools_results}.
For our approach, we partition the state space in $3,200$ rectangular regions.
It is fair to note that the safe set in SReachTools is smooth, while ours is not (due to our partition-based approach). 
While our current implementation is limited to regions as hyper-rectangles and parallelepipeds, our method can in theory be used with all convex polytopic regions.

\subsubsection{Reachability guarantees.}
We compare our method (with $N=25,\ldots,1600$ samples of the noise) to the results obtained via the different methods in \sreachtools. 
We only consider the particle and Voronoi methods of \sreachtools, because only these methods are sampling-based like our approach (although only the Voronoi method can give confidence guarantees on the results).
The reachability guarantees and run times are presented in \cref{tab:SReachTools}, and simulated trajectories in \cref{fig:SReachTools_results}.
As expected, our reachability guarantees, as well as for the Voronoi method, are a lower bound on the average simulated performance (and are thus safe), while this is not the case for the particle method of \sreachtools.

\subsubsection{Complexity and run time.}
As shown in \cref{tab:SReachTools}, the grid-free methods of \sreachtools are generally faster than our abstraction-based approach. 
However, we note that our method was designed for non-convex problems, which cannot be solved by \sreachtools.
Interestingly, the complexity of the particle~\citep{DBLP:conf/cdc/LesserOE13} and Voronoi partition method~\citep{DBLP:conf/amcc/SartipizadehVAO19} increases \emph{exponentially} with the number of samples, because their optimization problems have a binary variable for every particle.
By contrast, the complexity of our method increases only \emph{linearly} with the number of samples, because it suffices to count the number of samples in every region, as discussed in \cref{subsec:PracticalUse}.

\subsubsection{Controller type.}
The particle and Voronoi methods of \sreachtools synthesize \emph{open-loop} controllers, which means that they cannot act in response to observed disturbances of the state.
Open-loop controllers do not consider any feedback, making such controllers unsafe in settings with significant noise levels~\citep{aastrom2010feedback}.
By contrast, we derive \emph{piecewise linear feedback} controllers.
This structure is obtained, because our controllers are based on a state-based control policy that maps every continuous state within the same region to the same abstract action.

\subsubsection{Robustness against disturbances.}
To demonstrate the importance of feedback control, we test both methods under stronger disturbances, by increasing the covariance of the noise by a factor 10.
As shown in \cref{tab:SReachTools}, the particle method yields a low reachability probability (with the simulated performance being even lower), and the Voronoi method was not able to generate a solution at all.
By contrast, our method still provides safe lower bound guarantees, which become tighter for increasing sample sizes.
Our state-based controller is robust against the stronger noise, because it is able to act in response to observed disturbances, unlike the open-loop methods of \sreachtools that results in a larger error in the simulated state trajectories, as also shown in \cref{fig:SReachTools_results}.
    
\fi

\end{document}